\pdfoutput=1
\documentclass[11pt,a4paper]{article}
\usepackage{jheppub}

\RequirePackage[utf8]{inputenc}

\usepackage{amsfonts}
\usepackage{amssymb}
\usepackage{comment}
\usepackage{graphicx}
\usepackage{amsmath}
\usepackage{tabu}
\usepackage{dsfont}
\usepackage{float}
\usepackage{amsthm}
\usepackage{slashed}
\usepackage{setspace}
\usepackage[labelformat=simple]{subcaption}

\usepackage[boxsize=0.5em,aligntableaux=center]{ytableau}
\usepackage[utf8]{inputenc}
\usepackage{amsmath, amsthm, amssymb, amsfonts}
\usepackage[font=small, labelfont=bf]{caption}
\usepackage{amsmath, amsthm, amssymb, amsfonts}
\usepackage{multirow}
\usepackage{graphicx}
\usepackage{float}
\usepackage[numbers,sort&compress]{natbib}
\usepackage{jheppub}
\usepackage{enumerate}
\usepackage{amsmath}
\usepackage{amsfonts}
\usepackage{amssymb}
\usepackage{graphicx}
\usepackage{cancel}
\usepackage{physics}
\usepackage{braket}
\usepackage{empheq}
\usepackage{color}
\usepackage{hyperref}
\usepackage{float}
\restylefloat{table}
\restylefloat{figure}
\usepackage{graphicx}
\graphicspath{ {./images/} }

\def\e{\epsilon}

\newcommand{\rb}{\right)}
\newcommand{\lb}{\left(}
\newcommand{\rbs}{\right]}
\newcommand{\lbs}{\left[}
\newcommand{\nn}{\nonumber}

\newcommand{\nc}{\newcommand}
\nc{\beq}{\begin{equation}}
\nc{\eeq}{\end{equation}}
\nc{\bea}{\begin{eqnarray}}
\nc{\eea}{\end{eqnarray}}

\nc{\be}{\begin{equation}}
\nc{\ee}{\end{equation}}
\def\ov{\overline}
\def\vo{\mathcal{V}}

\nc{\at}{\bigg|}

\allowdisplaybreaks[2]
\numberwithin{equation}{section}

\def\V{\mathcal{V}}

\allowdisplaybreaks[2]
\numberwithin{equation}{section}

\nc{\ve}[1]{\left \langle #1 \right \rangle}

\title{Higher Derivative Corrections to String Inflation}

\author[1,2]{Michele Cicoli,}
\author[1,2]{Matteo Licheri,}
\author[1,2]{Pellegrino Piantadosi,}
\author[3,4]{Fernando Quevedo,}
\author[5]{Pramod Shukla}

\affiliation[1]{\footnotesize Dipartimento di Fisica e Astronomia, Universit\`a di Bologna, via Irnerio 46, 40126 Bologna, Italy}
\affiliation[2]{\footnotesize INFN, Sezione di Bologna, viale Berti Pichat 6/2, 40127 Bologna, Italy}
\affiliation[3]{\footnotesize DAMTP, Centre for Mathematical Sciences, Wilberforce Road, Cambridge, CB3 0WA, UK.}
\affiliation[4]{\footnotesize Perimeter Institute for Theoretical Physics, 31 Caroline Street North, Waterloo ON, Canada}
\affiliation[5]{\footnotesize Department of Physical Sciences, Unified Academic Campus, Bose Institute, \\ EN 80, Sector V, Bidhannagar, Kolkata 700 091, India}

\footnotesize\emailAdd{michele.cicoli@unibo.it} 
\footnotesize\emailAdd{matteo.licheri@unibo.it}
\footnotesize\emailAdd{ pellegrin.piantados2@unibo.it}
\footnotesize\emailAdd{f.quevedo@damtp.cam.ac.uk} 
\footnotesize\emailAdd{pshukla@jcbose.ac.in}

\abstract{We quantitatively estimate the leading higher derivative corrections to ${\mathcal{N}}=1$ supergravity derived from IIB string compactifications and study how they may affect moduli stabilisation and LVS inflation models. Using the Kreuzer-Skarke database of 4D reflexive polytopes and their triangulated Calabi-Yau database, we present scanning results for a set of divisor topologies corresponding to threefolds with $1 \leq h^{1,1} \leq 5$. In particular, we find several geometries suitable to realise blow-up inflation, fibre inflation and poly-instantons inflation, together with a classification of the divisors topologies for which the leading higher derivative corrections to the inflationary potential vanish. In all other cases, we instead estimate numerically how these corrections modify the inflationary dynamics, finding that that they do not destroy the predictions for the main cosmological observables.}

\keywords{Higher derivative corrections, Divisor topology, Blow-up inflation, Fibre inflation, Poly-instanton inflation}

\begin{document}

\maketitle

\bigskip

\section{Introduction}
\label{sec_intro}

Understanding supersymmetric effective field theories (EFT) from string compactifications is key in order to determine most of the relevant physical implications of these frameworks. These EFTs are only known approximately, and corrections to leading order effects play an important role for the most pressing questions such as moduli stabilisation and inflation from string theory.

These effects correspond to non-perturbative contributions to the superpotential $W$, and perturbative and non-perturbative corrections to the K\"ahler potential $K$, both in the $\alpha'$ and string-loop expansions. These corrections to $K$ and $W$ modify the standard $F$-term part of the scalar potential which comes from the square of the auxiliary fields at order $F^2$. However, there are also higher derivative $F^4$ corrections to the scalar potential. In the type IIB case, they have an explicit linear dependence on the two-cycle volume moduli $t^i$, $i=1,...,h^{1,1}$, and the overall volume $\mathcal{V}$ of the Calabi-Yau (CY) threefold $X$ \cite{Ciupke:2015msa,Cicoli:2016chb}:
\be
V_{F^4}=\frac{\gamma}{\vo^4}\,\sum_{i=1}^{h^{1,1}} \Pi_i t^i\,,
\label{VF4}
\ee
where $\gamma$ is a computable constant (independent of K\"ahler moduli) and $\Pi_i = \int_X c_2 \wedge \hat{D_i}$ with $c_2$ the CY second Chern-class and $\hat{D}_i$ a basis of harmonic (1,1)-forms dual to the divisors $D_i$. In terms of this basis, the K\"ahler form $J$ can be written as $J=t^i \hat{D}_i$.

The relevance of the corrections (\ref{VF4}) is manifest especially for determining the structure of the scalar potential since, due to the no-scale property, the leading order, tree-level, contribution vanishes, and therefore a combination of subleading corrections has to be considered. However, these higher-derivative corrections are naturally subdominant compared with the leading order $\alpha'^3$ correction at order $F^2$ that scales with the volume as $V_{\alpha'^3}\simeq |W_0|^2/\mathcal V^3$. In this sense they should not substantially modify moduli stabilisation mechanisms such as KKLT and the Large Volume Scenario (LVS). However, they can play a crucial role for:
\begin{enumerate}
\item Lifting flat directions which are not stabilised at leading LVS order \cite{Cicoli:2016chb};

\item Modifying slow-roll conditions needed for inflationary scenarios where the leading order effects leave an almost flat direction for the inflaton field.
\end{enumerate}
In this article we will concentrate on the second item, and find under which topological conditions these higher derivative corrections vanish. For cases where they are instead non-zero, we will numerically estimate the largest value of their prefactor $\gamma$ which does not ruin the flatness of the inflationary potential of different inflation models derived in the LVS framework such as blow-up inflation \cite{Conlon:2005jm,Blanco-Pillado:2009dmu,Cicoli:2017shd}, fibre inflation \cite{Cicoli:2008gp,Cicoli:2016xae, Burgess:2016owb,Cicoli:2017axo,Cicoli:2018cgu,Bhattacharya:2020gnk,Cicoli:2020bao,Cicoli:2022uqa} and poly-instanton inflation \cite{Cicoli:2011ct,Blumenhagen2012,Blumenhagen:2012ue,Lust:2013kt,Gao:2013hn}.

Using the Kreuzer-Skarke database of four-dimensional reflexive polytopes \cite{Kreuzer:2000xy} and their triangulated CY database \cite{Altman:2014bfa}, we present scanning results for a set of divisor topologies corresponding to CY threefolds with $1 \leq h^{1,1} \leq 5$.  These divisor topologies are relevant for various phenomenological purposes in LVS models. For inflationary model building, this includes, for example: ($i$) the (diagonal) del Pezzo divisors needed for generating non-perturbative superpotential corrections useful for blow-up inflation, ($ii$) the K3-fibration structure relevant for fibre inflation, and ($iii$) the so-called `Wilson' divisors which are relevant for realising poly-instanton inflation. In addition, we present a class of divisors which have vanishing $\Pi$.  

In this article we present general classes of divisor topologies which are relevant for making such corrections naturally vanish for the inflaton direction. In particular, we find that blow-up inflation is protected against such higher derivative corrections if the inflaton corresponds to the volume of a dP$_3$ divisor, i.e. a del Pezzo surface of degree six. Fibre inflation is instead shielded if the inflaton is the volume of a ${\mathbb T}^4$-divisor, while poly-instanton inflation is naturally safe only for the inflaton being the volume of a so-called `Wilson' divisor $(W)$, i.e. a rigid divisor with a Wilson line and $h^{1,1}(W)=2$. We present an explicit CY orientifold setting for each of these three classes of models. Moreover, we find that there are additional divisor topologies for which such $F^4$ corrections vanish. 

For generic topologies with non-vanishing $\Pi$, we perform a numerical estimate of the effect of these $F^4$ corrections on inflation, paying particular attention to the study of reheating from moduli decay to determine the exact number of efoldings of inflation which is relevant to match observations. We find that higher derivative $\alpha'^3$ effects do not substantially change the conclusions of fibre, blow-up and poly-instanton inflationary scenarios, therefore making those scenarios more robust under these corrections.

This article is organised as follows: Sec. \ref{sec_LVS} presents a brief review of LVS moduli stabilisation and the role of divisor topologies in LVS phenomenology. Subsequently we present a classification of the divisor topologies relevant for taming higher derivative $F^4$ corrections in Sec. \ref{sec_F4}. Sec. \ref{sec_BI} discusses instead potential candidate CYs for realising global embeddings of blow-up inflation and the effect of $F^4$ corrections on these models. The analysis of higher derivative corrections to LVS inflation models is continued in Sec. \ref{sec_FI} which is devoted to fibre inflation, and in Sec. \ref{sec_PI} which focuses on poly-instanton inflation. Finally, we summarise our results and present our conclusions in Sec. \ref{sec_conclusions}.

\section{Divisor topologies in LVS}
\label{sec_LVS}

In this section we present a brief review of the role of divisor topologies in the context of the LVS scheme of moduli stabilisation. It has been well established that some divisor topologies play a central role in LVS model building. These are, for example, del Pezzo (dP) and K3 surfaces. Such studies and suitable CY scans have been presented at several different occasions with different sets of interests \cite{Cicoli:2011it, Gao:2013pra, Altman:2014bfa, Cicoli:2018tcq, Cicoli:2021dhg,Altman:2021pyc,Gao:2021xbs,Carta:2022web,Shukla:2022dhz,Crino:2022zjk}, and we recollect some of the ingredients from \cite{AbdusSalam:2020ywo,Cicoli:2021dhg} which are relevant for the present work.

\subsection{Generic LVS scalar potential}

In the standard approach of moduli stabilisation in 4D type IIB effective supergravity models, one follows a so-called two-step strategy. In the first step, the axio-dilaton $S$ and the complex structure moduli $U^\alpha$ are stabilised by the superpotential $W_{\rm flux}$ induced by background 3-form fluxes $(F_3, H_3)$. This flux-dependent superpotential can fix all complex structure moduli and the axio-dilaton supersymmetrically at leading order by enforcing:
\be
D_{U^\alpha} W_{\rm flux} = D_S W_{\rm flux} \, = 0 \,.
\label{UStab}
\ee
After fixing $S$ and the $U$-moduli, the flux superpotential can effectively be considered as constant: $W_0=\langle W_{\rm flux}\rangle$. At this leading order, the K\"ahler moduli $T_i$ remain flat due to the no-scale cancellation. Using non-perturbative effects is one of the possibilities to fix these moduli. In this context, if we assume $n$ non-perturbative contributions to $W$ which can be generated by using rigid divisors, such as shrinkable dP 4-cycles, or by rigidifying non-rigid divisors using magnetic fluxes \cite{Bianchi:2011qh, Bianchi:2012pn, Louis:2012nb}, the superpotential takes the following form:
\be
W= W_0 + \sum_{i = 1}^n \, A_i\, e^{- a_i\, T_i}\,,
\label{eq:Wnp-n}
\ee
where:
\be
T_i = \tau_i + i \theta_i\qquad\text{with}\qquad \tau_i = \frac12 \int_{D_i} J\wedge J\quad\text{and}\quad \theta_i=\int_{D_i}\,C_4\,.
\ee
For the current work we consider CY orientifolds with trivial odd sector in the $(1,1)$-cohomology and subsequently orientifold-odd moduli are absent in our analysis (interested readers may refer to \cite{Gao:2014uha,Cicoli:2021tzt,Carta:2022web}). Note that in (\ref{eq:Wnp-n}) there is no sum in the exponents $(a_i\, T_i)$, and summations are to be understood only when upper indices are contracted with lower indices; otherwise we will write an explicit sum as in (\ref{eq:Wnp-n}). We will suppose that, out of $h^{1,1}_+ = h^{1,1}$ K\"ahler moduli, only the first $n$ appear in $W$, i.e. $i=1,...,n\leq h^{1,1}_+$. 

The K\"ahler potential including $\alpha'^3$ corrections takes the form \cite{Becker:2002nn}:
\be
\label{eq:K}
K = -\ln\left[-i\int \Omega\,(U^\alpha)\wedge\bar{\Omega}\,(\bar{U^\alpha})\right]-\ln\left(S+\bar{S}\right)-2\ln\left[{\cal V}\,(T_i+\bar{T_i})+\frac{\xi}{2}\left(\frac{S+\bar{S}}{2}\right)^{3/2}\right], \nonumber
\ee
where $\Omega$ denotes the nowhere vanishing holomorphic 3-form which depends on the complex-structure moduli, while ${\cal V}$ denotes the CY volume which receives $\alpha'^3$ corrections through $\xi=-\frac{\chi(X)\,\zeta(3)}{2\,(2\pi)^3}$ where $\chi(X)$ is the CY Euler characteristic and $\zeta(3)\simeq 1.202$.

Assuming that $S$ and the $U$-moduli are stabilised as in (\ref{UStab}), considering a superpotential given by (\ref{eq:Wnp-n}) and an $\alpha'^3$-corrected K\"ahler potential given by (\ref{eq:K}), one arrives at the following master formula for the scalar potential \cite{AbdusSalam:2020ywo}:
\be
V = V_{\alpha'^3} + V_{\rm np1} + V_{\rm np2} \,,
\label{eq:Vgen-nGen}
\ee
where (defining $\hat\xi\equiv \xi g_s^{-3/2}$ with $g_s=\langle{\rm Re}(S)\rangle^{-1}$):
\bea
\label{MasterF}
V_{\alpha'^3} &=&  e^K \, \frac{3 \, \hat\xi (\vo^2 + 7\,\vo\, \hat\xi +\hat\xi^2)}{({\cal V}-\hat{\xi }) (2\vo + \hat{\xi })^2}\, \,|W_0|^2\,, \\
V_{\rm np1} &=& e^K\, \sum_{i =1}^n \, 2 \, |W_0| \, |A_i|\, e^{- a_i \tau_i}\, \cos(a_i\, \theta_i + \phi_0 - \phi_i) \nn \\
&& \times ~\biggl[\frac{(4 \vo^2 + \vo \, \hat\xi+ 4\, \hat\xi^2)}{ (\vo - \hat\xi) (2\vo + \hat\xi)}\, (a_i\, \tau_i) +\frac{3 \, \hat\xi (\vo^2 + 7\,\vo\, \hat\xi + \hat\xi^2 )}{(\vo-\hat\xi) (2\vo + \hat\xi)^2}\biggr]\,, \nn \\
V_{\rm np2} &=& e^K\, \sum_{i=1}^n\,  \sum_{j=1}^n \, |A_i|\, |A_j| \, e^{-\, (a_i \tau_i + a_j \tau_j)} \, \cos(a_i\, \theta_i - a_j \, \theta_j -\phi_i + \phi_i) \,\nn \\
&& \times \biggl[ -4 \left({\cal V}+\frac{\hat\xi}{2}\right) \, (k_{ijk}\,t^k) \, a_i\, a_j\, + \frac{4{\cal V} - \hat{\xi}}{(\vo - \hat\xi)} \left(a_i\, \tau_i) \, (a_j\,\tau_j \right) \nn \\
&& + \,\frac{(4\vo^2  + \vo \, \hat{\xi} + 4\, \hat{\xi}^2)}{(\vo - \hat{\xi}) (2\vo + \hat{\xi})}\, (a_i\, \tau_i +a_j\, \tau_j) +\frac{3 \, \hat{\xi} (\vo^2 + 7\,\vo\, \hat{\xi}  +\hat\xi^2)}{({\cal V}-\hat{\xi }) (2\vo + \hat{\xi })^2}\biggr]~, \nn
\eea
where we have introduced phases into the parameters as $W_0=|W_0|\, e^{{\rm i} \, \phi_0}$ and $A_i = |A_i|\, e^{{\rm i}\, \phi_i}$. The good thing about the master formula (\ref{MasterF}) is the fact that it determines the complete form of $V$ simply by specifying topological quantities such as the intersection numbers $k_{ijk}$, the CY Euler number and the number $n$ of non-perturbative contributions to $W$.

Note that $V_{\alpha'^3}$ vanishes for $\hat\xi = 0$ and reproduces the standard no-scale structure in the absence of a $T$-dependent non-perturbative $W$. On the other hand, for very large volume $\vo \gg \hat\xi$, this term takes the standard form which plays a crucial r\^ole in LVS models \cite{Balasubramanian:2005zx}:
\be
V_{\alpha'^3} \simeq \left(\frac{g_s\,e^{K_{\rm cs}}}{2\,\vo^2}\, \right) \frac{3\,\hat\xi\, |W_0|^2}{4\, {\cal V}}\,.
\label{Valphaprime}
\ee
Let us also stress that $V_{\alpha'^3}$ depends only on the overall volume $\vo$, while $V_{\rm np1}$ depends on $\vo$ and the 4-cycle moduli $\tau_i$ (with the additional dependence on the axions $\theta_i$). Hence these two contributions to $V$ could be minimised by taking derivatives with respect to $\vo$ and $(h^{1,1}-1)$ 4-cycle moduli. However $V_{\rm np2}$ depends on the quantity $k_{ijk}\,t^k$ which in general cannot be inverted to be expressed as an explicit function of the $\tau_i$'s. It has been observed that using the master formula (\ref{MasterF}) one can efficiently perform moduli stabilisation in terms of the 2-cycle moduli $t^i$ as shown in \cite{AbdusSalam:2020ywo, AbdusSalam:2022krp}. 

For example, considering $h^{1,1}=2$, $n=1$ and $\hat\xi>0$ in the master formula (\ref{MasterF}) along with using the large volume limit, one can immediately read-off the following three terms:
\bea
\label{Vlvs}
V &\simeq& \frac{g_s\,e^{K_{\rm cs}}}{2} \biggl[ \frac{3 \, \hat\xi \, |W_0|^2}{4 \vo^3} + \frac{4 a_1 \tau_1 |W_0| |A_1|}{\vo^2}\, e^{- a_1 \tau_1} \cos\left(a_1 \theta_1 + \phi_0 - \phi_1\right) \\
&-& \frac{4 a_1^2 |A_1|^2 k_{111} t_1}{\vo}\, e^{-2 a_1 \tau_1} \biggr]. \nonumber
\eea
If the CY $X$ has a Swiss-cheese form, one can find a basis of divisors such that the only non-zero intersection numbers are $k_{111}$ and $k_{222}$. This leads to the relation $t^1 = - \sqrt{2\tau_1/k_{111}}$, where the minus sign is dictated from the K\"ahler cone conditions as the divisor $D_1$ in this Swiss-cheese CY is an exceptional 4-cycle. Using this in (\ref{Vlvs}) one gets \cite{Balasubramanian:2005zx}:\footnote{Ref. \cite{AbdusSalam:2020ywo} has shown that LVS moduli fixing can be realised also for generic cases where the CY threefold does not have a Swiss-cheese structure.}
\be
V \simeq \frac{g_s\, e^{K_{\rm cs}}}{2} \left( \frac{\beta_{\alpha'}}{\vo^3} + \beta_{\rm np1}\,\frac{\tau_1}{\vo^2}\, e^{- a_1 \tau_1} \cos\left(a_1 \theta_1 + \phi_0 - \phi_1\right) \\
+ \beta_{\rm np2}\,\frac{ \sqrt{\tau_1}}{\vo}\, e^{-2 a_1 \tau_1} \right), 
\label{VlvsSimpl}
\ee
with:
\be
\beta_{\alpha'} = \frac{3 \hat\xi |W_0|^2}{4}\,, \qquad \beta_{\rm np1} = 4 a_1 |W_0| |A_1|\,, \qquad \beta_{\rm np2} = 4 a_1^2 |A_1|^2 \sqrt{2 k_{111}}\,. 
\ee 

\subsection{Scanning results for LVS divisor topologies}

Let us start by briefly reviewing the generic methodology for analysing the divisor topologies which is widely adopted for scanning useful CY geometries suitable for phenomenology, e.g. see \cite{Cicoli:2021dhg,Shukla:2022dhz}. Subsequently we will continue following the same in our current analysis. The main idea is to consider the CY threefolds arising from the four-dimensional reflexive polytopes listed in the Kreuzer-Skarke (KS) database \cite{Kreuzer:2000xy}, and classify the divisors based on their relevance for phenomenological model building aiming at explicit orientifold constructions. For that purpose, we rather have a very nice collection of the various topological data of CY threefolds available in the database of \cite{Altman:2014bfa} which can be directly used for further analysis. In this regard, Tab. \ref{tab_PTG} presents the number of (favorable) polytopes along with the corresponding (favorable) triangulations and (favorable) geometries for a given $h^{1,1}(X)$ in the range $1 \le h^{1,1}(X) \le 5$.

\begin{table}[h!]
  \centering
 \begin{tabular}{|c|c|c||c|c||c|c||}
\hline
 $h^{1,1}$ & Polytopes  & Favorable & Triangs.  & Favorable   & Geometries & Favorable   \\
& & Polytopes & & Triangs. & &   Geoms.    \\
 \hline
 1 & 5 & 5 & 5 & 5 & 5  & 5  \\
 2 & 36 & 36 & 48 & 48 & 39  & 39  \\
 3 & 244 & 243 & 569 & 568 & 306  & 305  \\
 4 & 1197 & 1185 & 5398 & 5380 & 2014  & 2000  \\
 5 & 4990 & 4897 & 57132 & 56796 & 13635  & 13494  \\
 \hline
  \end{tabular}
\caption{Number of (favourable) triangulations and (favourable) distinct CY geometries arising from the (favourable) polytopes listed in the Kreuzer-Skarke database.}
\label{tab_PTG}
\end{table}

For a given CY geometry, the main focus is limited to:
\begin{itemize}
\item 
looking at the topology of the so-called `coordinate divisors' $D_i$ which are defined through setting the toric coordinates to zero, i.e. $x_i = 0$. This means that there is a possibility of missing a huge number of divisors, e.g. those which could arise via considering some linear combinations of the coordinate divisors, and some of such may have interesting properties. However, it is hard to make an exhaustive analysis including all the effective divisors of a given CY threefold.

\item
focusing on scans using `favourable' triangulations (Triang$^*$) and `favourable' geometries (Geom$^*$) for a given polytope. This could be justified in the sense that for non-favourable CY threefolds, the number of toric divisors in the basis is less than $h^{1,1}(X)$, and subsequently there is always at least one coordinate divisor which is non-smooth, and one usually excludes such spaces from the scan. However, the number of such CY geometries is almost negligible in the sense that there are just 1, 14 and 141 for $h^{1,1}(X)$ being 3, 4 and 5 respectively.
\end{itemize} 

The role of divisor topologies in the LVS context can be appreciated by noting that the Swiss-cheese structure of the CY volume can be correlated with the presence of del Pezzo (dP$_n$) divisors $D_s$. These dP$_n$ divisors are defined for $ 0 \leq n \leq 8$ having degree $d = 9 -n$ and $h^{1,1} = 1 + n$, such that dP$_0$ is a ${\mathbb P}^2$ and the remaining 8 del Pezzo's are obtained by blowing up eight generic points inside ${\mathbb P}^2$. It turns out that they satisfy the following two conditions \cite{Cicoli:2011it}:
\be
\label{eq:dP}
\int_X D_s^3 = k_{sss} > 0\, , \qquad \int_X D_s^2 \, D_i \leq 0 \qquad \forall \, i \neq s \,.
\ee
Here the self-triple-intersection number $k_{sss}$ corresponds to the degree of the del Pezzo 4-cycle dP$_n$ where $k_{sss} = 9 - n > 0$, which is always positive as $n \leq 8$ for del Pezzo surfaces. In addition, one imposes the so-called `diagonality' condition on such a del Pezzo divisor $D_s$ via the following relation satisfied by the triple intersection numbers \cite{Cicoli:2011it, Cicoli:2018tcq}:
\be
\label{eq:diagdP}
k_{sss} \, \, k_{s i j } = k_{ss i} \, \, k_{ss j} \, \qquad \qquad \forall \, \, \, i, j.
\ee
It turns out that whenever this diagonality condition is satisfied, there exists a basis of coordinates divisors such that the volume of each of the 4-cycles $D_s$ becomes a complete-square quantity as illustrated from the following relations:
\be
\tau_s = \frac{1}{2}\, k_{s ij} t^i \, t^j = \frac{1}{2 \, k_{sss}}\, k_{ssi} \, k_{s s j} t^i \, t^j = \frac{1}{2 \, k_{sss}}\, \left(k_{ss i} \,t^i \, \right)^2\,.
\ee
Subsequently what happens is that one can always shrink such a `diagonal' del Pezzo ddP$_n$ to a point-like singularity by squeezing it along a single direction. A systematic analysis on counting the CY geometries which could support (standard) LVS models, in the sense of having at least one diagonal del Pezzo divisor, has been performed in \cite{Cicoli:2021dhg} and the results are summarised in Tab. \ref{tab_ddPns-GstarM}. Moreover, it is worth mentioning that the scanning result presented in Tab. \ref{tab_ddPns-GstarM} is quite peculiar in the sense that for all the CY threefolds with $h^{1,1} \leq 5$, one does not have any example having a `diagonal' dP$_n$ divisor for $ 1 \leq n \leq 5$, which has been subsequently conjectured to be true for all the CY geometries arising from the KS database. 

\noindent
\begin{table}[H]
	\centering
	\hskip0.11cm \begin{tabular}{|c||c|c||c|c|c|c|c|c||c|}
		\hline
		$h^{1,1}$ & Poly$^*$ &  Geom$^*$ & $\mathrm{ddP}_0$  & $d{\mathbb F}_0$  & $\mathrm{ddP}_n$ & $\mathrm{ddP}_6$  & $\mathrm{ddP}_7$  & $\mathrm{ddP}_8$  & $n_{\rm LVS}$ \\
		&  & ($n_{\rm CY}$) &  &  & $1\leq n \leq 5$ &  &  &  & ($\mathrm{ddP}_n\geq 1$) \\
		\hline
		1 & 5 & 5 & 0 & 0  & 0  & 0 & 0 & 0 & 0 \\
		2 & 36 & 39 & 9 & 2 & 0 & 2  & 4  & 5 & 22 \\
		3 & 243 & 305 & 59 & 16 & 0  & 17 & 40 & 39 & 132\\
		4 & 1185 & 2000 & 372 & 144 & 0 & 109 & 277 & 157 &  750 \\
		5 & 4897 & 13494 & 2410 & 944  & 0  & 624 & 827 & 407 & 4104 \\
		\hline
	\end{tabular}
	\caption{Number of CY geometries with a `diagonal' del Pezzo divisor suitable for LVS. Here we have extended the notation to denote a ${\mathbb P}^2$ surface as ddP$_0$ and a diagonal ${\mathbb P}^1 \times {\mathbb P}^1$ surface as $d{\mathbb F}_0$.}
	\label{tab_ddPns-GstarM}
\end{table}

Let us mention that the classification of CY geometries relevant for LVS as presented in Tab. \ref{tab_ddPns-GstarM} corresponds to having a `standard' LVS in the sense of having at least one `diagonal' del Pezzo divisor in a Swiss-cheese like model. However, it has been found in some cases that one can still have alternative moduli stabilisation schemes realising an exponentially large CY volume, e.g. using the underlying symmetries of the CY threefold in the presence of a non-diagonal del Pezzo \cite{AbdusSalam:2020ywo}, and in the framework of the so-called perturbative LVS \cite{Antoniadis:2018hqy,Antoniadis:2019rkh, Leontaris:2022rzj,Leontaris:2023obe}.

\section{Topological taming of $F^4$ corrections}
\label{sec_F4}

In addition to the $\alpha'^3$ correction (\ref{Valphaprime}) derived in \cite{Becker:2002nn}, generically there can be many other perturbative corrections to the 4D effective scalar potential induced from various sources (see \cite{Burgess:2020qsc,Cicoli:2021rub} for a classification of potential contributions at different orders in $\alpha'$ exploiting higher dimensional rescaling symmetries and F-theory techniques). One such effect are $F^4$ corrections which cannot be captured by the two-derivative ansatz for the K\"ahler and superpotentials. In this section we shall discuss the topological taming of such corrections in the context of LVS inflationary model building.

\subsection{$F^4$ corrections to the scalar potential}

The higher derivative $F^4$ contributions to the scalar potential for a generic CY orientifold compactification take the following simple form \cite{Ciupke:2015msa}:
\be
\label{VF4orig}
V_{F^4} = - \left(\frac{e^{K_{cs}}\, g_s}{8 \pi}\right)^2 \frac{\lambda\,|W_0|^4}{g_s^{3/2} {\cal V}^4} \sum_{i=1}^{h^{1,1}}\Pi_i \, t^i \, \equiv \frac{\gamma}{{\cal V}^4} \, \sum_{i=1}^{h^{1,1}}\ \Pi_i\, t^i,
\ee
where the topological quantities $\Pi_i$ are given by:
\be
\label{Pii}
\Pi_i = \int_X c_2(X) \wedge \hat{D}_i \,,
\ee
and $\lambda$ is an unknown combinatorial factor which in the single modulus case is rather small in absolute value \cite{Grimm:2017okk}:
\be
\lambda = -\frac{11}{24} \, \frac{\zeta(3)}{(2 \pi)^4} = -3.5 \cdot 10^{-4}\,.
\label{eq:suppresion}
\ee
Its value is not known for $h^{1,1}>1$ but we expect it to remain small, in analogy with the $h^{1,1}=1$ case. In fact, one can argue that the factor $\zeta(3)/(2\pi)^4$ in $\lambda$ is expected to be always present for generic models with several K\"ahler moduli as well. This is because the coupling tensor ${\cal T}_{\ov{i}\,\ov{j}\, k l}$ appearing in this correction through the following higher derivative piece \cite{Ciupke:2015msa}:
\be
V_{F^4} = - e^{2 K} \, {\cal T}^{\ov{i}\,\ov{j} k l} \ov{D}_{\ov i} \ov{W}\, \ov{D}_{\ov j} \ov{W}\,D_k {W}\,{D}_{l} {W}\,,
\label{VgenF4}
\ee
can be schematically written as:
\be
{\cal T}_{\ov{i}\,\ov{j}\, k l} =\frac{c}{\mathcal{V}^{8/3}} \, \frac{\zeta(3)\, {\cal Z}}{g_s^{3/2}}\,,
\ee
where $c$ can be considered as some combinatorial factor, which for example, in the single modulus case turns out to be $11/384$ \cite{Grimm:2017okk}, and:
\be
{\cal Z} = (2 \pi)^2\, \int_X c_2(X) \wedge J \,,
\ee
where we stress that we are working with the convention $\ell_s = (2 \pi) \sqrt{\alpha'} = 1$. Subsequently, we have
\be
{\cal T}_{\ov{i}\,\ov{j}\, k l} = c \, \frac{\zeta(3)}{(2\pi)^{4}\,\mathcal{V}^{8/3}\, g_s^{3/2}} \int_X c_2(X) \wedge J\,.
\ee
Note that the $\mathcal{V}^{-8/3}$ factor in the above expression cancels off with a $\mathcal{V}^{8/3}$ contribution coming from 4 inverse K\"ahler metric factors needed to raise the 4 indices of the coupling tensor ${\cal T}_{\ov{i}\,\ov{j}\, k l}$ to go to (\ref{VgenF4}).

Here, let us mention that the higher derivative $F^4$ correction under consideration appears at $\alpha'^3$ order, like the BBHL-correction \cite{Becker:2002nn}, and both are induced at string tree-level, resulting in a factor of $g_s^{-3/2}$. For explicitness, let us also note that the leading order BBHL correction \cite{Becker:2002nn} appearing at the two-derivative level takes the following form:\footnote{In this regard, it may be worth noticing that the original result \cite{Becker:2002nn} has been obtained with the convention $(2\pi \alpha') = 1$ which removes the $(2\pi)^{-3}$ factor from the denominator of the $\hat\xi$ parameter.}
\be
 V_{\alpha'^3} = \left(\frac{e^{K_{cs}}\, g_s}{8 \pi}\right) \, \frac{3\, \xi\,|W_0|^2}{4\,g_s^{3/2}\,{\cal V}^3}, \qquad \xi = -\frac{\zeta(3) \chi(X)}{2\, (2\pi)^3}\,.
\ee
Now, comparing these two $\alpha'$ corrections one finds that:
\be
\label{eq:ratio-VF4-F2}
\frac{V_{F^4}}{V_{\alpha'^3}} = \tilde{c}\,\left(\frac{g_s}{8 \pi}\right) \, e^{K_{cs}}\,|W_0|^2 \, \left(\frac{\Pi_i \, t^i}{\chi(X){\cal V}}\right), 
\ee
where $\tilde{c}$ is some combinatorial factor, which for the case of a single K\"ahler modulus is
\be
\tilde{c} = \frac{11}{9(2\pi)} \simeq 0.2\,.
\ee
One can observe that each factors in (\ref{eq:ratio-VF4-F2}) can be of a magnitude less than one in typical models. For example, demanding large complex-structure limit in order to ignore instanton effects can typically result in having $e^{K_{cs}} \sim 0.01$ \cite{Louis:2012nb}, the string coupling $g_s$ needs to be small and the CY volume large to trust the low-energy EFT, and the ratios between $\Pi_i$'s and $\chi(X)$ are typically of ${\cal O}(1)$ \cite{Shukla:2022dhz}. Having these aspects in mind, it is very natural to anticipate that higher-derivative $F^4$ effects are subdominant as compared to the two-derivative corrections. Note that (\ref{VF4orig}) can also be rewritten as:
\begin{equation}
  V_{F^4}  = - V_{\alpha'^3}\, \frac{\sqrt{g_s}}{3\pi}\left(\frac{\lambda}{\xi}\right)\sum_{i=1}^{h^{1,1}}\Pi_i\left(\frac{m_{3/2}}{M_{\rm KK}^{(i)}}\right)^2
  \label{VF4new}
\end{equation}
where the gravitino mass is:
\begin{equation}
    m_{3/2}^2 = \left(\frac{g_s}{8\pi}\right) \frac{|W_0|^2}{\vo^2}\, M_p^2\,,
\end{equation}
and $M_{\rm KK}^{(i)}$ is the Kaluza-Klein scale associated to the $i$-th divisor:
\begin{equation}
    \left(M_{\rm KK}^{(i)}\right)^2 = \frac{M_s^2}{t_i} = \frac{\sqrt{g_s}}{4\pi}\frac{M_p^2}{t_i\vo}\,.
\end{equation}
In the above equation we have used the relation between the string scale and the Planck mass in the convention where $\mathcal{V}_s=\mathcal{V}\,g_s^{3/2}$ (with $\mathcal{V}_s$ the volume in string frame and $\mathcal{V}$ the volume in Einstein frame):
\begin{equation}
    M_s^2 = \frac{1}{(2\pi)^2 \alpha'}=\sqrt{g_s} \frac{M_p^2}{4\pi \vo}\,.
\end{equation}
Note that (\ref{VF4new}) makes clear that $V_{F^4}$ is an $\mathcal{O}(F^4)$ correction since $V_{\alpha'^3}$ is an $\mathcal{O}(F^2)$ effect and \cite{Cicoli:2013swa}:
\begin{equation}
\left(\frac{m_{3/2}}{M_{\rm KK}}\right)^2 \sim g \frac{|F|^2}{M_{\rm KK}^2}\ll 1\,,
\end{equation}
where $g\sim M_{\rm KK}/M_p \sim \vo^{-2/3}\ll 1$ is the coupling of heavy KK modes to light states.

\subsection{Classifying divisors with vanishing $F^4$ terms}

Two important quantities characterising the topology of a divisor $D$ are the Euler characteristic $\chi(D)$ and the holomorphic Euler characteristic (also known as arithmetic genus) $\chi_h(D)$ which are given by the following useful relations \cite{Blumenhagen:2008zz, Collinucci:2008sq, Cicoli:2016xae}:
\bea
\label{chih}
\chi(D) &\equiv& \sum_{i=0}^4 {(-1)}^i \, b_i(D)  =  \int_X \, \hat{D} \wedge\left(\hat{D} \wedge \hat{D} + c_2(X) \right)\,, \label{chi} \\
\chi_h({D}) &\equiv& \sum_{i=0}^2 {(-1)}^i \, h^{i,0}(D)  = \frac{1}{12} \int_X \, \hat{D} \wedge \left(2 \, \hat{D}\wedge \hat{D} + c_2(X) \right)\,, 
\eea
where $b_i(D)$ and $h^{i,0}(D)$ are respectively the Betti and Hodge numbers of the divisor. Using these two relations we find that $\Pi(D)$ is related with the Euler characteristics and the holomorphic Euler characteristic as follows:
\be
\label{PiAndChis}
\Pi(D) = \chi(D) -  \int_X \, \hat{D} \wedge \hat{D} \wedge \hat{D}, \qquad  \Pi(D) = 12 \, \chi_h(D) -  2 \, \int_X \, \hat{D} \wedge \hat{D} \wedge \hat{D} \, ,
\ee
which also give another useful relation:
\be
\Pi(D) = 2 \, \chi(D) - 12 \, \chi_h(D) \, .
\ee
Therefore, the topological quantity $\Pi(D)$ vanishes for a generic smooth divisor $D$ if the following simple relation holds,
\be
\label{eq:PiZero0}
\Pi(D) = 0  \quad \Longleftrightarrow  \quad \chi(D) = 6 \, \chi_h(D) \,.
\ee
Now, using the relations $\chi(D) =2 \, h^{0,0} - 4 \, h^{1,0} + 2\, h^{2,0} + h^{1,1}$ and $\chi_h(D) = h^{0,0} - h^{1,0}+h^{2,0}$, we find another equivalent relation for vanishing $\Pi(D)$:
\be
\label{PiZero1}
h^{1,1}(D) = 4 \, h^{0,0}(D) - 2\, h^{1,0}(D) + 4 \, h^{2,0}(D) \,.
\ee
Any divisor satisfying the vanishing $\Pi$ relation (\ref{PiZero1}) will be denoted as $D_\Pi$. After knowing the topology of a generic divisor $D$, it is easy to check if $h^{1,1}$ satisfies this condition or equivalently $\chi = 6\, \chi_h$. To demonstrate it, let us quickly consider the following two examples:
\bea
{\mathbb T}^4 \equiv
\begin{tabular}{ccccc}
    & & 1 & & \\
   & 2 & & 2 & \\
  1 & & 4 & & 1 \\
   & 2 & & 2 & \\
    & & 1 & & \\
  \end{tabular} \qquad \qquad\text{and}\qquad\qquad
	{\rm K3} \equiv
\begin{tabular}{ccccc}
    & & 1 & & \\
   & 0 & & 0 & \\
  1 & & 20 & & 1 \\
   & 0 & & 0 & \\
    & & 1 & & \\	
  \end{tabular} \, . \nonumber
	\eea
Now it is obvious that ${\mathbb T}^4$ has $\Pi({\mathbb T}^4)=0$ as it satisfies $\chi = 0 = 6\, \chi_h$. However, K3 has $\Pi({\rm K3})=24$ and $6 \chi_h = 12 = \chi/2$. Alternatively, it can be also checked that the Hodge number condition in (\ref{PiZero1}) is satisfied for ${\mathbb T}^4$ but not for K3. 

Therefore, we can generically formulate that a divisor $D$ of a Calabi-Yau threefold having the following Hodge Diamond results in a vanishing $\Pi(D)$:
\bea
\label{eq:DPi0}
& & D_{\Pi} \equiv
\begin{tabular}{ccccc}
    & & $h^{0,0}$ & & \\
   & $h^{1,0}$ & & $h^{1,0}$ & \\
  $h^{2,0}$ \quad \, \, & & $\left(4 h^{0,0}- 2 h^{1,0}+ 4 h^{2,0}\right)$ & & \quad $h^{2,0}$ \\
   & $h^{1,0}$ & & $h^{1,0}$ & \\
    & & $h^{0,0}$ & & \\
  \end{tabular}\, ,
  \eea
and if we consider that the $D_\Pi$ divisor is smooth and connected, then we have $h^{0,0}(D_\Pi) = 1$. Subsequently we can identify three different classes of vanishing $\Pi$ divisors:
\begin{enumerate}
\item{\textbf{dP$_3$ divisors:} For connected rigid 4-cycles with no Wilson lines we have $h^{1,0}(D) = h^{2,0}(D)=0$, and hence a vanishing $\Pi(D)$ results in the following Hodge diamond:
\bea
\label{eq:DPi}
& & D_\Pi \equiv
\begin{tabular}{ccccc}
    & & 1 & & \\
   & 0 & & 0 & \\
0 & & 4 & & 0 \\
   & 0 & & 0 & \\
    & & 1 & & \\
\end{tabular} \equiv {\rm dP}_{\Pi}\,.
\eea
This topology corresponds to the dP surface of degree six, i.e. a dP$_3$. Moreover, this class of $D_\Pi$ which singles out a dP$_3$ surface, also includes the possibility of the `rigid but not del Pezzo' 4-cycle denoted as NdP$_n$ for $ n \geq 9$ \cite{Cicoli:2011it}. These surfaces are blow-up of line-like singularities and  have similar Hodge diamonds as those of the usual dP surfaces dP$_n$ defined for $0 \leq n \leq 8$.}

\item{\textbf{Wilson divisors:} For connected rigid 4-cycles with Wilson lines we have $h^{2,0}(D) = 0$ but $h^{1,0}(D) > 0$, resulting in the following Hodge diamond for $D_{\Pi}$:
\bea
\label{eq:DPiw}
& & D_{\Pi} \equiv
\begin{tabular}{ccccc}
    & & 1 & & \\
   & $h^{1,0}$ & & $h^{1,0}$ & \\
  0 \quad & & $\left(4 - 2 h^{1,0}\right)$ & & \quad 0 \\
   & $h^{1,0}$ & & $h^{1,0}$ & \\
    & & 1 & & \\
\end{tabular} \equiv W_\Pi\,.
\eea
Given that all Hodge numbers are non-negative integers, the only possibility compatible with $h^{1,1}\geq 1$ (to be able to a have a proper definition of the divisor volume) is $h^{1,0} =1$ which, in turn, corresponds to $h^{1,1}=2$. This is a so-called `Wilson' divisor with vanishing $\Pi(W)$ which we denote as $W_\Pi$. This $W_\Pi$ divisor corresponds to a subclass of `Wilson' divisors, characterised by the Hodge numbers $h^{0,0} = h^{1,0}=1$ and arbitrary $h^{1,1}$, that have been introduced in \cite{Blumenhagen:2012kz} to support poly-instanton corrections.}

\item{\textbf{Non-rigid divisors:} Now let us consider the third special class which can have deformation divisors, i.e. $h^{2,0}(D) > 0$. When the divisor does not admit any Wilson line, i.e. $h^{1,0}(D) =0$, the Hodge diamond for $D_{\Pi}$ simplifies to:
\bea
\label{eq:DPi}
& & D_{\Pi} \equiv
\begin{tabular}{ccccc}
    & & 1 & & \\
   & 0 & & 0 & \\
  $h^{2,0}$ \, \, \, & & $\left(4+ 4 \, h^{2,0}\right)$ & & \qquad $h^{2,0}$ \\
   & 0 & & 0 \\
    & & 1 & & \\
\end{tabular}\,.
\eea
To our knowledge, so far there are no known examples in the literature which have such a topology. The simplest of its kind will have $h^{2,0}(D) =1$ and $h^{1,1}(D) = 8$. In this regard, it is worth mentioning that the topology of the so-called `Wilson' divisors which are ${\mathbb P}^1$ fibrations over ${\mathbb T}^2$s, has been argued to be useful in \cite{Lust:2006zg} and some years later it was found to be the case while studying the generation of poly-instanton effects \cite{Blumenhagen:2012kz}. So it would be interesting to know if such non-rigid divisor topologies of vanishing $\Pi$ exist in explicit CY constructions, and further if they could be useful for some phenomenological applications.

The last possibility is to consider the most general situation with deformations and Wilson lines, i.e. $h^{2,0}(D) > 0$ and $h^{1,0}(D) > 0$. As already mentioned, the simplest case is $\mathbb{T}^4$ with $h^{2,0}(\mathbb{T}^4) = 1$, $h^{1,0}(\mathbb{T}^4) =2$ and $h^{1,0}(\mathbb{T}^4) =4$ which however never shows up in our search through the KS list, as well as more general divisors with both deformations and Wilson lines.}
\end{enumerate}

Before coming to the scan of such divisor topologies of vanishing $\Pi$, let us mention a theorem of \cite{Oguiso1993, Schulz:2004tt} which states that if the CY intersection polynomial is linear in the homology class $\hat{D}_f$ corresponding to a divisor $D_f$, then the CY threefold has the structure of a K3 or a ${\mathbb T}^4$ fibration over a ${\mathbb P}^1$ base. Noting the following relation for the self-triple-intersection number of a generic smooth divisor $D$:
\be
\label{eq:chisAndDDD}
\int_X \hat{D} \wedge \hat{D} \wedge \hat{D} = 12 \, \chi_h(D) - \chi(D) \,,
\ee
and subsequently demanding the absence of such cubics for $D_f$ in the CY intersection polynomial, results in $\chi(D) = 12\chi_h(D)$ or the following equivalent relation: 
\be
\label{eq:cubicZero}
\quad h^{1,1}(D_f) = 10\, h^{0, 0}(D_f) - 8\, h^{1,0}(D_f) +10\, h^{2,0}(D_f)\,.
\ee
This relation is clearly satisfied for K3 and ${\mathbb T}^4$ divisors, and can be satisfied for some other possible topologies as well. For example, another non-rigid divisor for which the self-cubic-intersection is zero is given by the following Hodge diamond:
\bea
& & {\rm SD} \equiv
\begin{tabular}{ccccc}
    & & 1 & & \\
   & 0 & & 0 & \\
  2 & & 30 & & 2 \\
   & 0 & & 0 & \\
    & & 1 & & \\	
  \end{tabular}, \qquad \chi(SD) = 36, \qquad \chi_h(SD) = 3 \, .\nonumber
	\eea
This is also a very well known surface frequently appearing in CY threefolds, e.g. it appears in the famous Swiss-cheese CY threefold defined as a degree-18 hypersurface in WCP$^4[1,1,1,6,9]$ where the divisors corresponding to the first three coordinates with charge 1 are such surfaces. 

Moreover, interestingly one can see that for the `Wilson' type divisor the relation in (\ref{eq:cubicZero}) is indeed satisfied for $h^{1,1}(D) =2$ which is exactly something needed for the generation of poly-instanton effects on top of having vanishing $\Pi(D)$ as we have discussed before. In this regard, let us also add that the simultaneous vanishing of $\Pi(D)$ and $D^3_{|_X}$ results in the vanishing of $\chi(D)$ and $\chi_h(D)$ and vice-versa, and so, besides a particular type of `Wilson' divisor, there can be more such divisor topologies satisfying the following if and only if condition:
\be
\label{eq:PiAndCubicZero}
\Pi(D) = 0  = \int_X\, \hat{D} \wedge \hat{D} \wedge \hat{D}  \qquad \Longleftrightarrow \qquad \chi(D) = 0 = \chi_h(D)\,.
\ee
Thus, if a divisor $D$ is connected and has $\Pi(D) = 0 = D^3_{|_X}$, then its Hodge diamond is:
\bea
\label{eq:D3PibothZero}
D_\Pi \equiv \begin{tabular}{ccccc}
    & & 1 & & \\
   & $1+n$ & & $1+n$ & \\
  $n$ \qquad & & $2+2 n$ & & \quad $n$ \\
   & $1+ n$ & & $1+n$ & \\
    & & 1 & & \\	
  \end{tabular}\equiv D_\Pi^{\rm cubic} \,,
 \eea
where $n$ is the number of possible deformations for the divisor $D$. For $n=0$ this corresponds to a $W_\Pi$ divisor, and for $n=1$ this corresponds to a ${\mathbb T}^4$. Although we are not aware of any such examples with $n \geq 2$, it would be interesting to know what topology they would correspond to.

\subsection{Scan for divisors with vanishing $F^4$ terms}

In this section we discuss the scanning results for divisors with $\Pi=0$ using the favorable CY geometries arising from the four-dimensional reflexive polytopes of the KS database \cite{Kreuzer:2000xy} and its pheno-friendly collection in \cite{Altman:2014bfa}. As pointed out earlier, we will consider only the `coordinate divisors' and the `favourable' CY geometries listed in Tab. \ref{tab_PTG}. For finding divisors with vanishing $\Pi$, we consider the following two different strategies in our scan:
\begin{enumerate}
\item{One route is to directly compute $\Pi$ by using the second Chern class of the CY threefold and the intersection tensor available in the database \cite{Altman:2014bfa}.}

\item{A second route is to compute the divisor topology using cohomCalg \cite{Blumenhagen:2010pv, Blumenhagen:2011xn} and subsequently to check the Hodge number condition (\ref{PiZero1}), or the equivalent relation $\chi(D) = 6 \, \chi_h(D)$, for vanishing $\Pi$.}
\end{enumerate}

Tab. \ref{tab_GallPi0LVS} presents the scanning results for the number of CY geometries with vanishing $\Pi$ divisors, and their suitability for realising LVS models. On the other hand, Tab. \ref{tab_GdP3Pi0LVS} and \ref{tab_GwilsonPi0LVS} show the same results split for the cases where the divisors with $\Pi=0$ are respectively dP$_\Pi$ (i.e. dP$_3$) and Wilson divisors $W_\Pi$. These distinct CY geometries and their scanning results correspond to the favourable geometries arising from the favourable polytopes.

\begin{table}[H]
  \centering
 \begin{tabular}{|c||c|c||c|c|c|c||c|c|c|}
\hline
 $h^{1,1}$ & Poly$^*$  & Geom$^*$ & single & two & three & four  & $n_{\rm LVS}$  & $n_{\rm LVS}$  & $n_{\rm LVS}$       \\
&  & $(n_{\rm CY})$  & $D_\Pi$ & $D_\Pi$ & $D_\Pi$  & $D_\Pi$  & \& 1 $D_\Pi$ & \& 2 $D_\Pi$ & \& 3 $D_\Pi$      \\
 \hline
 1 & 5 & 5 & 0 & 0  & 0 & 0 & 0  & 0 & 0   \\
 2 & 36 & 39 & 0 & 0  & 0 & 0 & 0  & 0 & 0   \\
 3 & 243 & 305 & 23  & 0  & 0  & 0  &  4 & 0 & 0    \\
 4 & 1185 & 2000 & 322 & 24  & 0  & 0  & 78  & 1 & 0  \\
 5 & 4897 &13494 & 3306  & 495  & 27 & 1 &  732 & 104  &  1  \\
 \hline
\end{tabular}
  \caption{CY geometries with vanishing $\Pi$ divisors and a ddP$_n$ to support LVS.}
  \label{tab_GallPi0LVS}
 \end{table}

\begin{table}[H]
  \centering
 \begin{tabular}{|c||c|c||c|c|c|c||c|c|c|c|}
\hline
  $h^{1,1}$ & Poly$^*$  & Geom$^*$  & At least  & Single & Two & Three  & $n_{\rm LVS}$ & $n_{\rm LVS}$ & $n_{\rm LVS}$    \\
&  & $(n_{CY})$ & one dP$_\Pi$ & dP$_\Pi$ & dP$_\Pi$  & dP$_\Pi$  & \& dP$_\Pi$  & \& 1 dP$_\Pi$  & \& 2 dP$_\Pi$     \\
 \hline
 1 & 5 & 5 & 0 & 0  & 0 & 0 & 0  & 0 & 0   \\
 2 & 36 & 39 & 0 & 0 & 0 & 0 & 0  & 0 &  0 \\
 3 & 243 & 305 & 4 &  4 & 0 & 0 & 0 & 0 &  0  \\
 4 & 1185 & 2000 & 143 &  134 & 9 & 0 & 16  & 16 &  0   \\
 5 & 4897 &13494 & 2236  & 2035  & 197 & 4 &  336 & 290  &  46 \\
 \hline
  \end{tabular}
  \caption{CY geometries with vanishing $\Pi$ divisors of the type dP$_\Pi\equiv$ dP$_3$, and a ddP$_n$ for LVS.}
   \label{tab_GdP3Pi0LVS}
 \end{table}
 \noindent

\begin{table}[H]
  \centering
 \begin{tabular}{|c||c|c||c|c|c|c||c|c|c|c|}
\hline
$h^{1,1}$ & Poly$^*$  & Geom$^*$  & At least  & Single & Two & Three  & $n_{\rm LVS}$ \&  & $n_{\rm LVS}$ \& & $n_{\rm LVS}$ \&     \\
&  & $(n_{CY})$ & one $W_\Pi$ & $W_\Pi$ & $W_\Pi$  & $W_\Pi$  & 1 $W_\Pi$  & 2 $W_\Pi$  & 3 $W_\Pi$     \\
 \hline
 1 & 5 & 5 & 0 & 0  & 0 & 0 & 0  & 0 & 0   \\
 2 & 36 & 39 & 0 & 0 & 0 & 0 & 0  & 0 &  0 \\
 3 & 243 & 305 & 19 & 19 & 0 & 0 & 4 & 0 &  0  \\
 4 & 1185 & 2000 & 210 & 202 & 8 & 0 & 62  & 1 &  0   \\
 5 & 4897 &13494 & 1764  & 1599  & 154 & 11 &  442 & 79  &  1 \\
 \hline
  \end{tabular}
  \caption{CY geometries with vanishing $\Pi$ divisors of the type $W_\Pi$, and a ddP$_n$ to support LVS.}
   \label{tab_GwilsonPi0LVS}
 \end{table}
 
To appreciate the scanning results presented in Tab. \ref{tab_GallPi0LVS}, \ref{tab_GdP3Pi0LVS} and \ref{tab_GwilsonPi0LVS} corresponding to all CY threefolds with $ 1 \leq h^{1,1}(X) \leq 5$ in the KS database, let us make the following generic observations:
\begin{itemize}

\item
We do not find any CY threefold in the KS database which has a ${\mathbb T}^4$ divisor or any divisor with vanishing $\Pi(D)$ and $h^{2,0}(D) \neq 0$. The only possible vanishing $\Pi$ divisors we encountered in our scan are either a dP$_3$ divisor or a Wilson divisor with $h^{1,1}(W)= 2$. However, going beyond the coordinate divisors in an extended scan as compared to ours may have more possibilities.

\item
For $h^{1,1}(X) =1$ and $2$, there are no CY threefolds with a vanishing $\Pi$ divisor.

\item
Although there are some dP$_3$ divisors for CY threefolds with $h^{1,1}(X) =3, 4$ and $5$, none of them are diagonal in the sense of being shrinkable to a point by squeezing along a single direction \cite{Cicoli:2018tcq} -- something in line with the conjecture of \cite{Cicoli:2021dhg}.

\item
There are no CY threefolds with $h^{1,1}(X) =3$ which have (at least) one diagonal dP$_n$ and a (non-diagonal) dP$_3$ with $\Pi({\rm dP}_3)=0$. Hence, in order to have a dP$_3$ divisor in LVS, we need CY threefolds with $h^{1,1}(X) \geq 4$. For $h^{1,1}(X) = 4$ there are 16 CY threefolds in the `favourable' geometry which are suitable for LVS and feature a dP$_3$.

\item
For $h^{1,1}(X) \leq 4$, there is only one CY geometry which can lead to LVS and has two vanishing $\Pi$ divisors which are of Wilson-type. Similarly, there is only one CY geometry with a ddP for LVS and 3 vanishing $\Pi$ divisors.
\end{itemize}

\section{Blow-up inflation with $F^4$ corrections}
\label{sec_BI}

The minimal LVS scheme of moduli stabilisation fixes the CY volume $\vo$ along with a small modulus $\tau_s$ controlling the volume of an exceptional del Pezzo divisor. Therefore any LVS model with 3 or more K\"ahler moduli, $h^{1,1}\geq 3$, can generically have flat directions at leading order. These flat directions are promising inflaton candidates with a potential generated at subleading order. Blow-up inflation \cite{Conlon:2005jm} corresponds to the case where the inflationary potential is generated by non-perturbative superpotential contributions. In this inflationary scenario the inflaton is a (diagonal) del Pezzo divisor wrapped by an ED3-instanton or supporting gaugino condensation. In addition, the CY has to feature at least one additional ddP$_n$ divisor to realise LVS. 

On these lines, we present the scanning results in Tab. \ref{tab_blowup-Gstar} corresponding to the number of CY geometries $n_{\rm CY}$ with their suitability for realising LVS along with resulting in the standard blow-up inflationary potential, in the sense of having at least two ddP divisors, one needed for supporting LVS and the other one for driving inflation.

\begin{table}[H]
\centering
\hskip0.11cm \begin{tabular}{|c|c|c||c|c|c|c||c|c|}
\hline
 $h^{1,1}$ & Poly$^*$ &  Geom$^*$ & $n_{\rm ddP}=1$  & $n_{\rm ddP}=2$ & $n_{\rm ddP}=3$  & $n_{\rm ddP}=4$ & $n_{\rm LVS}$ & Blow-up \\
&  & $(n_{\rm CY})$ &  &   &  &  &  & infl. \\
 \hline
 1 & 5 & 5 & 0 & 0  & 0  & 0 & 0 & 0 \\
 2 & 36 & 39 & 22 & 0  & 0  & 0 & 22 & 0 \\
 3 & 243 & 305 & 93 & 39  & 0 & 0 & 132 & 39 \\
 4 & 1185 & 2000 & 465 & 261 & 24 & 0 & 750 & 285 \\
 5 & 4897 & 13494 & 3128  & 857  & 106 & 13 & 4104 & 976 \\
 \hline
  \end{tabular}
\caption{Number of LVS CY geometries suitable for blow-up inflation.}
   \label{tab_blowup-Gstar}
 \end{table}

\subsection{Inflationary potential}

The simplest blow-up inflation model is based on a two-hole Swiss-cheese CY threefold. Such a CY threefold has two diagonal del Pezzo divisors, say $D_1$ and $D_2$, which after considering an appropriate basis of divisors result in the following intersection polynomial:
\be
I_3 = I_3^\prime(D_{i^\prime}) + k_{111} \, D_{1}^3 + k_{222} \, D_{2}^3 \,, \quad {\rm for} \, \, {i^\prime} \neq \{1, 2\}\,,
\ee
where $I_3^\prime(D_{i^\prime})$ is such that $D_1$ and $D_2$ do not appear in this cubic polynomial. Further, $k_{111}$ and $k_{222}$ are the self-intersection numbers which are fixed by the degrees of the two del Pezzo divisors, say dP$_{n_1}$ and dP$_{n_2}$, as $k_{111} = 9 - n_1 >0$ and  $k_{222} = 9 - n_2 >0$. This generically provides the following expression for the volume form:
\be
{\cal V} = \frac{1}{6}\, k_{i'j'k'} \, t^{i'} \, t^{j'}\, t^{k'} + \frac{k_{111}}{6} \, (t^1)^3 + \frac{k_{222}}{6} \, (t^2)^3 \, ,
\ee
where the 2-cycle volume moduli $t^{i^\prime}$ are such that $i' \neq \{1, 2\}$. Subsequently, the volume can be rewritten in terms of the 4-cycle volume moduli as:
\be
{\cal V} = f_{3/2}(\tau_{i'}) - \beta_1 \, \, \tau_{1}^{3/2} - \beta_2\, \, \tau_{2}^{3/2} \, ,
\ee
where $\beta_1 = \frac{1}{3} \sqrt{\frac{2}{k_{111}}}$ and $\beta_2 = \frac{1}{3} \sqrt{\frac{2}{k_{222}}}$. Furthermore, under our choice of the intersection polynomial, $\tau_{i'}$ does not depend on the del Pezzo volumes $\tau_1$ and $\tau_2$. Now we can simplify things to the minimal three-field case with $h^{1,1}_+ = 3$ by taking $f_{3/2}(t^{i'}) = \frac{1}{6}\, k_{bbb}\, (t^b)^3$ and using the following relations between the 2-cycle moduli $t^i$ and the 4-cycle moduli $\tau_i$:
\be
\label{eq:t2tau-BI}
t^b = \, \sqrt{\frac{2\,\tau_{b}}{k_{bbb}}}\,, \qquad t^1 = - \, \sqrt{\frac{2\,\tau_{1}}{k_{111}}} \,, \qquad t^2 = - \, \sqrt{\frac{2\,\tau_{2}}{k_{222}}} \,.
\ee

The scalar potential of the minimal blow-up inflationary model \cite{Conlon:2005jm, Cicoli:2017shd} can be reproduced by the master formula (\ref{MasterF}) via simply setting $h^{1,1}_+=3$, $n=2$ and $\hat\xi>0$, which leads to the following leading order terms in the large volume limit:
\bea 
\label{eqn:BLOWUPpotential}
V &=& \frac{e^{K_{\rm cs}}}{2 s} \left[ \frac{3 \hat\xi |W_0|^2}{4\vo^3} + \sum_{i=1}^2 \frac{4 |W_0| |A_i| a_i}{\vo^2}\,\tau_i\, e^{- a_i \tau_i}\, \cos(a_i \theta_i + \phi_0 - \phi_i)\right. \label{3LVS} \\
&-& \left. \sum_{i=1}^2 \sum_{j=1}^2 \frac{4 |A_i| |A_j| a_i a_j}{\vo} \, e^{- (a_i\tau_i + a_j\tau_j)} \, \cos(a_j \theta_j - a_i \theta_i -\phi_j + \phi_i) \left(\sum_{k=1}^3 k_{ijk} t_k \right)\right]. \nn 
\eea
Given that we are interested in a strong Swiss-cheese case where the only non-vanishing intersection numbers are $k_{111}$, $k_{222}$ and $k_{333}$, we have:
\be
\sum_{k=1}^3 k_{iik}t_k = k_{iii} t_i = - \sqrt{2 \, k_{iii}\, \tau_i} \quad\text{for}\,\,i=1,2\qquad\text{and}\qquad \sum_{k=1}^3 k_{ijk}t_k = 0\quad\text{for}\,\,i\neq j\,. \nn
\ee
Hence (\ref{3LVS}) reduces to the potential of known 3-moduli Swiss-cheese LVS models \cite{Conlon:2005jm, Cicoli:2017shd}:
\be
V = \frac{e^{K_{\rm cs}}}{2 s} \left[ \frac{\beta_{\alpha'}}{\vo^3} + \sum_{i=1}^2 \left(\beta_{{\rm np1},i}\,\frac{\tau_i}{\vo^2}\, e^{- a_i \tau_i}\, \cos(a_i \theta_i + \phi_0 - \phi_i) + \beta_{{\rm np2},i}\,\frac{\sqrt{\tau_i}}{\vo} \, e^{- 2 a_i\tau_i}\right) \right], \nn 
\ee
with:
\be
\beta_{\alpha'} = \frac{3 \hat\xi |W_0|^2}{4}\,, \qquad \beta_{{\rm np1},i} = 4 a_1 |W_0| |A_1|\,, \qquad \beta_{{\rm np2},i} = 4 a_1^2 |A_1|^2 \sqrt{2 k_{111}}\,. 
\ee
It has been found that such a scalar potential can drive inflation effectively by a single field after two moduli are stabilised at their respective minimum \cite{Conlon:2005jm}. In fact, a three-field inflationary analysis has been also presented in \cite{Blanco-Pillado:2009dmu,Cicoli:2017shd} ensuring that one can indeed have trajectories which effectively correspond to a single field dynamics.
 
\subsection{$F^4$ corrections}

In this three-field blow-up inflation model, higher derivative $F^4$ corrections to the scalar potential look like:\footnote{Additional perturbative corrections can arise from string loops but we assume that these contributions can be made negligible by either taking a small value of the string coupling or by appropriately small flux-dependent coefficients.}
\bea
\label{eq:VF4-BI}
V_{F^4} &= & \frac{\gamma}{{\cal V}^4} \, \left(\Pi_b\, t^b  + \Pi_1 \, t^1 + \, \Pi_2 \, t^2 \right)\, \\
& = & \frac{\gamma}{{\cal V}^4} \, \left(\Pi_b\, t^b  - \Pi_1 \, \sqrt{\frac{2\,\tau_{1}}{k_{111}}}  - \, \Pi_2 \, \sqrt{\frac{2\,\tau_{2}}{k_{222}}} \right) \nonumber\\
&  = & \frac{\gamma}{{\cal V}^4} \, \left(\Pi_b\,  \left(\frac{6}{k_{bbb}}\right)^{1/3}\, \left({\cal V} +\beta_1 \, \, \tau_{1}^{3/2} + \beta_2\, \, \tau_{2}^{3/2} \right)^{1/3}  - \Pi_1 \, \sqrt{\frac{2\,\tau_{1}}{k_{111}}}  - \, \Pi_2 \, \sqrt{\frac{2\,\tau_{2}}{k_{222}}} \right), \nonumber
\eea
where we have used the relations in (\ref{eq:t2tau-BI}). Assuming that inflation is driven by $\tau_2$, only $\tau_2$-dependent corrections can spoil the flatness of the inflationary potential.
The leading correction is proportional to $\Pi_2$ and scales as ${\cal V}^{-4}$, while a subdominant contribution proportional to $\Pi_b$ would scale as ${\cal V}^{-14/3}$. It is interesting to note that this subleading correction would be present even if $\Pi_2=0$, as in the case where the corresponding dP$_n$ is a diagonal dP$_3$. As compared to the LVS potential, this inflaton-dependent $F^4$ correction is suppressed by a factor of order ${\cal V}^{-5/3}\ll 1$. Moreover, the ideal situation to completely nullify higher derivative $F^4$ corrections for blow-up inflation is to demand that:
\be
\Pi_b = \Pi_2 = 0\,.
\ee
In this setting, making $\Pi_b$ zero by construction appears to be hard and very unlikely since we have seen that vanishing $\Pi$ divisors other than dP$_3$ could possibly be either a ${\mathbb T}^4$ or a Wilson divisor. However, for both divisors we have $\int_X\, D^3 = 0$ as they satisfy the condition (\ref{eq:cubicZero}) that implies vanishing cubic self-intersections, and so they do not seem suitable to reproduce the strong Swiss-cheese volume form that has been implicitly assumed in rewriting the scalar potential pieces in (\ref{eq:VF4-BI}). Moreover, we have not observed any other kind of vanishing $\Pi$ divisors in our scan involving the whole set of CY threefolds with $h^{1,1} \leq 5$ in the KS database.\footnote{However recall that our scan is limited to coordinate divisors only, and so may miss some possibilities.} Let us finally point out that a case with $\Pi_b=0$ cannot be entirely ruled out as we have seen in a couple of non-generic situations that a non-fibred K3 surface can also appear as a `big' divisor in a couple of strong Swiss-cheese CY threefolds, and so if there is a similar situation in which a non-fibred ${\mathbb T}^4$ appears with a ddP divisor it could possibly make $\Pi_b$ identically zero.

\subsection{Constraints on inflation}

We are now going to study the effect of $F^4$ corrections in blow-up inflation, focusing on the case where their coefficients are in general non-zero, as suggested by our scan. In this analysis we shall follow the work of \cite{Cicoli2016a}. First of all, we will derive the value of the volume to subsequently analyse the effect of the $F^4$ corrections to the inflationary dynamics. 

We start from the potential described in \eqref{eqn:BLOWUPpotential}, stabilise the axions and set $e^{K_{\text{cs}}}/(2s)=1$, obtaining:
\be\label{eqn:LVS_potential}
    V_{\text{LVS}} = \sum_{i=1}^{2} \left( \frac{8(a_i A_i)^2\sqrt{\tau_i}}{3\vo \beta_i} e^{-2 a_i \tau_i}-\frac{4 a_i A_i W_0\tau_i}{\vo^2}e^{-a_i \tau_i} \right) + \frac{3\hat{\xi}W_0^2}{4\vo^3} \, ,
\ee
where the volume has been expressed as:
\be 
    \vo = \tau_b^{3/2} - \beta_1 \tau_1^{3/2} -\beta_2 \tau_2^{3/2} \, .
\ee
The minimum condition of the LVS potential reads:
\be
    e^{-a_i \tau_i} = \frac{\Lambda_i}{\vo}\sqrt{\tau_i} \,,
    \label{eqn:LVS_minimum_condition}
\ee
where the constants $\Lambda_i$ are defined as:
\be
    \Lambda_i \equiv \frac{3 |W_0|}{4} \frac{\beta_i}{a_i |A_i|} \,.
\ee
Moreover, since we want to find an approximate Minkowski vacuum, we add an uplifting potential of the generic form:
\be
V_{\text{up}} = \frac{D}{\vo^{4/3}}\,,
\ee
where the value of $D$ will be computed in the next paragraph. Lastly, the $F^4$ corrections become:
\be
V_{F^4} = \frac{\gamma}{\vo^4} \left[ 
\Pi_b \left(\vo -  \sum_{i=1}^{2}\beta_i \tau_i^{3/2} \right)^{1/3} - 3 \sum_{i=1}^2 \Pi_i\beta_i \sqrt{\tau_i} \right].
\ee

\subsubsection{Volume after inflation}

We start by fixing in the LVS potential \eqref{eqn:LVS_potential} the small moduli at their minimum given by \eqref{eqn:LVS_minimum_condition}:
\be
V_{\text{LVS}} = -\frac{3 |W_0|^2}{2\vo^3} \lb \sum_{i=1}^{2}\beta_i\tau_i^{3/2}
    -\frac{\hat{\xi}}{2} \rb \,.
\ee
Defining $\psi \equiv \ln\vo$, the minimum condition for $\tau_i$ can be approximated as:
\be
    \tau_i = \frac{1}{a_i} \lb \psi -\ln \Lambda_i -\ln\sqrt{\tau_i} \rb \simeq \frac{1}{a_i} \lb \psi -\ln \Lambda_i \rb \, ,
\ee
leading to:
\be
V_{\text{LVS}}^{({\rm PI})} = -\frac{3 |W_0|^2}{4} e^{-3\psi} \lbs \sum_{i=1}^{2}P_i\lb
	\psi- \ln\Lambda_i \rb^{3/2}-\hat{\xi} \rbs \, ,
\ee
where $P_i \equiv 2\beta_i a_i^{-3/2}$ and the superscript $({\rm PI})$ indicates that we consider the `post inflation' situation where all the moduli reach their minimum. Analogously, the uplifting term reads:
\be
V_{\text{up}}^{({\rm PI})} = D e^{-\frac43\psi} \,,
\ee
while the $F^4$ correction becomes:
\be
V_{F^4}^{({\rm PI})} = \gamma e^{-4\psi} \lbs 
\Pi_b \lb e^\psi - \sum_{i=1}^{2}P_i\lb
\psi- \ln\Lambda_i \rb^{3/2} \rb^{1/3} - 3 \sum_{i=1}^{2}\Pi_i P_i a_i \lb
\psi- \ln\Lambda_i \rb^{1/2} \rbs \,.
\ee
The full post-inflationary potential for the field $\psi$ is therefore:
\be
V_{\text{PI}}(\psi) = V_{\text{LVS}}^{({\rm PI})}+V_{\text{up}}^{({\rm PI})}+V_{F^4}^{({\rm PI})}\,.
\ee
We are now able to calculate the factor $D$ in order to have a Minkowski minimum, by imposing:
\be
V_{\text{PI}}^{\prime}(\tilde{\psi}) = V_{\text{PI}}(\tilde{\psi}) = 0\, ,
\ee
which gives:
\begin{align}
D =&\, \frac{27 |W_0|^2}{20} e^{-\frac53\tilde{\psi}} \sum_i P_i \lb\tilde{\psi}-\ln\Lambda_i\rb^{1/2}
	+\delta D_{F^4} \, ,\\
	\delta D_{F^4} =&\, \gamma e^{-\frac83\tilde{\psi}} \Bigg[ 
    \Pi_b \lb      e^{\tilde{\psi}} -
        \sum_{i}P_i\lb \tilde{\psi}- \ln\Lambda_i \rb^{3/2} 
\rb^{1/3} - 3 \sum_{i}\Pi_i P_i a_i \lb \tilde\psi- \ln\Lambda_i \rb^{1/2} \nonumber \\
&- \frac{\Pi_b}{3} \frac{
e^{\tilde{\psi}}-\frac32 \sum_i P_i \lb \tilde{\psi}- \ln\Lambda_i \rb^{1/2}}{
        \lb e^{\tilde{\psi}} - \sum_{i}P_i\lb \tilde{\psi}- \ln\Lambda_i \rb^{3/2} \rb^{2/3}} +     \frac32 \sum_{i} \Pi_i P_i a_i \lb \tilde{\psi} - \ln\Lambda_i \rb^{-1/2}
	\Bigg] \, ,
\end{align}
where $\tilde{\psi}$ solves the following equation:
\begin{align}
\sum_i P_i &\lb \tilde\psi -\ln\Lambda_i \rb^{1/2}\lb \tilde\psi -\ln\Lambda_i -\frac{9}{10}\rb-\hat\xi + \delta\tilde\psi_{F^4} = 0\, , \\
\delta\tilde\psi_{F^4} =&\, -\frac{4\gamma}{5W_0^2} e^{-\tilde{\psi}} \Bigg[ \frac{8\Pi_b}{3} \lb 
e^{\tilde{\psi}} - \sum_{i}P_i\lb \tilde{\psi}- \ln\Lambda_i \rb^{3/2} \rb^{1/3} -
8 \sum_{i}\Pi_i P_i a_i \lb \tilde\psi- \ln\Lambda_i \rb^{1/2} \nonumber \\
&- \frac{\Pi_b}{3} \frac{
e^{\tilde{\psi}}-\frac32 \sum_i P_i \lb \tilde{\psi}- \ln\Lambda_i \rb^{1/2}}{\lb e^{\tilde{\psi}} - \sum_{i}P_i\lb \tilde{\psi}- \ln\Lambda_i \rb^{3/2} \rb^{2/3} } + 
    \frac32 \sum_{i} \Pi_i P_i a_i \lb \tilde{\psi} - \ln\Lambda_i \rb^{-1/2}	\Bigg] \, ,
\end{align}
from which we obtain the post inflation volume $\vo_{\text{PI}} \equiv e^{\tilde\psi}$.

\subsubsection{Volume during inflation}

We now move on to determine the value of the volume modulus during inflation. In order to do so, we focus on the region in field space where the inflaton $\tau_2$ is away from its minimum. In this region, the inflaton-dependent contribution to the volume potential becomes negligible due to the large exponential suppression from \eqref{eqn:LVS_potential}. Hence, the inflationary potential for the volume mode is given only by:
\be
V_{\text{inf}}(\psi) = -\frac{3 |W_0|^2}{4} e^{-3\psi} \lbs P_1\lb
\psi- \ln\Lambda_1 \rb^{3/2}-\hat{\xi} \rbs + D e^{-\frac43\psi} \, ,
\ee
where we ignore $F^4$ corrections since the volume during inflation is bigger than the post-inflationary one. At this point we can again minimise the $\psi$ field to a value $\hat{\psi}$, imposing the vanishing of the first derivative:
\be 
P_1 \lb \hat\psi -\ln\Lambda_1 \rb^{1/2}\lbs 2\lb \hat\psi -\ln\Lambda_1\rb -1\rbs-2\hat\xi + \frac{16}{9|W_0|^2} D e^{\hat\psi}  = 0  \, ,
\ee
and the volume during inflation is given as $\vo_{\text{inf}} \equiv \e^\psi$.

\subsubsection{Inflationary dynamics}

During inflation all the moduli, except $\tau_2$, sit at their minimum, including the volume mode which is located at $\vo\equiv \vo_{\text{inf}}$. From now on, we will drop the subscript and always refer to the volume as the one during inflation, unless otherwise explicitly stated. The inflaton potential with higher derivative effects reads:
\begin{align}
V(\tau_2) =&\, V_0 - \frac{4 a_2 |A_2| |W_0| }{\V^2}\tau_2 e^{-a_2 \tau_2} + \frac{8\,a_2^2\, |A_2|^2\sqrt{\tau_2}}{3\vo \beta_2} e^{-2 a_2 \tau_2} \\ &+ \frac{\gamma}{\vo^4} \left[ 
\Pi_b \left(\vo - P_1  \lb \ln\frac{\V}{\Lambda_1} \rb^{3/2}- \beta_2 \tau_2^{3/2} \right)^{1/3} - 3  \sum_i\Pi_i\beta_i \sqrt{\tau_i} \right]. \nonumber
\end{align}
Canonically normalising the inflaton field as:
\be
\tau_2 = \lb \ve{\tau_2}^{3/4}+\sqrt{\frac{3 \V}{4 \beta_2}}\,\phi\rb^{4/3}\,,
\ee
we find the inflaton effective potential:
\begin{align}
V(\phi) =&\, V_0 - \frac{4 a_2 |A_2| |W_0| }{\V^2} \lb \sqrt{\frac{3 \V}{4 \beta_2}} \phi+\ve{\tau_2}^{3/4}\rb^{4/3} e^{-a_2 \lb \sqrt{\frac{3 \V}{4 \beta_2}}\phi+\ve{\tau_2}^{3/4}\rb^{4/3} } \nn \\
	&+ \frac{8\,a_2^2\, |A_2|^2}{3\vo \beta_2}\lb \sqrt{\frac{3 \V}{4 \beta_2}}\phi+\ve{\tau_2}^{3/4}\rb^{2/3} e^{-2 a_2 \lb \sqrt{\frac{3 \V}{4 \beta_2}}\phi+\ve{\tau_2}^{3/4}\rb^{4/3}} \nn \\
	&+ \frac{\gamma}{\vo^4} \bigg[
        \Pi_b \left(\vo - P_1  \lb \ln\frac{\V}{\Lambda_1} \rb^{3/2}- \beta_2 \lb \sqrt{\frac{3 \V}{4 \beta_2}}\phi+\ve{\tau_2}^{3/4}\rb^{2} \right)^{1/3} \nn\\
    &- 3  \Pi_2\beta_2 \lb \sqrt{\frac{3 \V}{4 \beta_2}}\phi+\ve{\tau_2}^{3/4}\rb^{2/3} - 3 \Pi_1 P_1 a_1 \lb \ln\frac{\V}{\Lambda_1} \rb^{1/2} 
    \bigg] \, .
\end{align}
To simplify the notation, we introduce:
\begin{gather}
A \equiv \frac{4 a_2 |A_2|\, |W_0|}{\V^2}\, ,\qquad 
B \equiv \frac{8\,a_2^2 |A_2|^2}{3\vo \beta_2} \, , \qquad C \equiv \V -P_1\lb \ln \frac{\V}{\Lambda_1} \rb^{3/2} \,,\\
\gamma_2 \equiv \frac{3\gamma\Pi_2\beta_2}{\V^4}\, , \qquad \gamma_b \equiv \frac{\gamma\Pi_b}{\V^4} \, ,\qquad \alpha \equiv \sqrt{\frac{3\V}{4\beta_2}} \,,\\
\varphi \equiv \sqrt{\frac{3 \V}{4 \lambda_2}}\phi+\ve{\tau_2}^{3/4} = \alpha \phi + \ve{\tau_2}^{3/4} \,,
\end{gather}
and we absorb the constant $F^4$ correction proportional to $\Pi_1$ inside $V_0 $ as:
\be
V_0 \to V_0 - \frac{3\gamma \Pi_1 P_1 a_1}{\V^4} \lb \ln\frac{\V}{\Lambda_1} \rb^{1/2}  \,.
\ee
The potential therefore simplifies to:
\begin{equation}
V(\varphi) = V_0 - A\,\varphi^{4/3} \,e^{-a_2\,\varphi^{4/3}} + B \varphi^{3/2} e^{-2 a_2\varphi^{4/3}} + \gamma_b (C-\beta_2\,\varphi^2)^{1/3} - \gamma_2 \varphi^{2/3} \,.
\end{equation} 
Given that $\varphi$ is different from the canonically normalised inflaton $\phi$, we define the following notation for differentiation:
\be
f'(\varphi) \equiv \dv{f(\varphi)}{\phi} = \sqrt{\frac{3 \V}{4 \lambda_2}}\dv{f(\varphi)}{\varphi} \equiv \alpha \dot{f}(\varphi)\,,
\ee
with the slow-roll parameters calculated as follows:
\begin{equation}
\epsilon = \frac{1}{2}\left(\frac{V'}{V}\right)^2 = \frac12\alpha^2\left(\frac{\dot{V}}{V}\right)^2 \qquad\text{and}\qquad
\eta = \frac{V''}{V} = \alpha^2 \frac{\ddot{V}}{V} \,.
\end{equation}
The next step is to find the value of $\phi$ at the end of inflation, which we denote as $\phi_{\text{end}}$, where $\epsilon(\phi_{\text{end}}) = 1$. Moreover, the number of efoldings from horizon exit to the end of inflation can be computed as:
\be
N_e(\phi_{\rm exit}) = \int_{\phi_{\text{end}}}^{\phi_{\rm exit}} \frac{\dd{\phi}}{\sqrt{2\epsilon}} = \int_{\varphi_{\text{end}}}^{\varphi_{\rm exit}} \frac{\dd{\varphi}}{\alpha\sqrt{2\epsilon}} \, .
\label{NeEq}
\ee
This value has to match the number of efoldings of inflation $N_e$ computed from the study of the post-inflationary evolution which we will perform in the next section, i.e. $\phi_{\rm exit}$ is fixed by requiring $N_e(\phi_{\text{exit}}) = N_e$. The observed amplitude of the density perturbations has to be matched at $\phi_{\rm exit}$, which typically fixes $\vo\sim 10^{5-6}$. The predictions for the main cosmological observable are then be inferred as follows:
\begin{equation}
n_s = 1+ 2 \eta(\phi_{\text{exit}}) - 6\epsilon(\phi_{\text{exit}})
\qquad\text{and}\qquad r = 16\epsilon(\phi_{\text{exit}})\,.
\end{equation}

\subsubsection{Reheating}

In order to make predictions that can be confronted with actual data, we need to derive the number of efoldings of inflation which, in turn, are determined by the dynamics of the reheating epoch. Assuming that the Standard Model is realised on a stack of D7-branes, a crucial term in the low-energy Lagrangian to understand reheating is the loop-enhanced coupling of the volume mode to the Standard Model Higgs $h$ which reads \cite{Cicoli:2022fzy}:
\be 
\mathcal{L} \subset c_{\text{loop}} \frac{m_{3/2}^2}{M_p}\,\phi_b h^2 \,,
\label{eqn:HiggsCouplingEnhanced}
\ee
where $c_{\rm loop}$ is a 1-loop factor and $\phi_b$ the canonically normalised volume modulus. Two different scenarios for reheating can arise depending on the presence or absence of a stack of D7-branes wrapped around the inflaton del Pezzo divisor:
\begin{itemize}
\item \textbf{No D7s wrapped around the inflaton:} The inflaton $\tau_2$ is not wrapped by any D7 stack and the Standard Model is realised on D7-branes wrapped around the blow-up mode $\tau_1$. This case has been studied in \cite{Cicoli:2022fzy}. The volume mode, despite being the lightest modulus, decays before the inflaton due to the loop-enhanced coupling (\ref{eqn:HiggsCouplingEnhanced}). Reheating is therefore caused by the decay of the inflaton which occurs with a width: 
\be
\Gamma_{\tau_2} \simeq \frac{1}{\V} \frac{m_{\tau_2}^3}{M_p^2}\simeq \frac{M_p}{\V^4} \,,
\ee
leading to a matter dominated epoch after inflation which lasts for the following number of efoldings:
\be
    N_{\tau_2} = \frac{2}{3} \ln\lb \frac{ H_{\text{inf}} }{\Gamma_{\tau_2}}\rb = \frac53 \ln\V \,.
    \ee
Thus, the total number of efoldings for inflation is given by:
\be
N_e = 57 + \frac14 \ln r - \frac14 N_{\tau_2} \simeq 50 - \frac14 N_{\tau_2} = 50 - \frac{5}{12}\ln\V\,,
\label{NeNoD7}
\ee
where we have focused on typical values of the tensor-to-scalar ratio for blow-up inflation around $r\sim 10^{-10}$. Thus, due to the long epoch of inflaton domination before reheating, the total number of required efoldings can be considerably reduced, resulting in a potential tension with the observed value of the spectral index, as we will point out in the next section. Note that the inflaton decay into bulk axions can lead to an overproduction of dark radiation which is however avoided by the large inflaton decay width into Standard Model gauge bosons, resulting in $\Delta N_{\rm eff}\simeq 0.13$ \cite{Cicoli:2022fzy}.
    
\item \textbf{D7s wrapped around the inflaton:} The inflaton is wrapped by a D7 stack which can be either the Standard Model or a hidden sector. These different cases have been analysed in \cite{Cicoli:2010ha, Cicoli:2010yj, Allahverdi:2020uax}. The localisation of gauge degrees of freedom on the inflaton divisor increases the inflaton decay width, so that the last modulus to decay is the volume mode. However the naive estimate of the number of efoldings of the matter epoch dominated by the oscillation of $\vo$ is reduced due to the enhanced Higgs coupling \eqref{eqn:HiggsCouplingEnhanced}. The early universe history is then given by a first matter dominated epoch driven by the inflaton which features now an enhanced decay rate:
\be
\Gamma_{\tau_2} \simeq \V \,\frac{m_{\tau_2}^3}{M_p^2} \simeq \frac{M_p}{\V^2} \,.
\ee
Hence the number of efoldings of inflaton domination is given by:
\be
N_{\tau_2} = \frac{2}{3} \ln\lb \frac{ H_{\text{inf}} }{\Gamma_{\tau_2}}\rb = \frac13 \ln\V \, .
\ee
The volume mode starts oscillating during the inflaton dominated epoch. Redshifting both as matter, the ratio of the energy densities of the inflaton and the volume mode remains constant from the start of the volume oscillations to the inflaton decay:
\be
\theta^2 \equiv \frac{\rho_{\tau_b}}{\rho_{\tau_2}}\at_{\text{osc}} = \frac{\rho_{\tau_b}}{\rho_{\tau_2}}\at_{\text{dec},\tau_2}\ll 1 \,,
\ee
since the energy density after inflation is dominated by the inflaton. Assuming that the inflaton dumps all its energy into radiation when it decays, we can estimate:
\be
\frac{\rho_{\tau_b}}{\rho_\gamma}\at_{\text{dec}} = \theta^2\,.
\ee
The radiation dominated era after the inflaton decay ends when $\rho_\gamma$ becomes comparable to $\rho_{\tau_b}$, which occurs when: \be
\rho_\gamma\at_{\text{dec},\tau_2} \left(\frac{a_{\text{dec},\tau_2}}{a_{\text{eq}}}\right)^4 =
     \rho_{\tau_b}\at_{\text{dec},\tau_2} \left(\frac{a_{\text{dec},\tau_2}}{a_{\text{eq}}}\right)^3 
     \quad\Rightarrow\quad a_{\text{dec},\tau_2} = a_{\text{eq}}\, \theta^2 \,,
    \ee
giving the dilution at equality:
\be
\rho_{\text{eq}} = \rho_{\gamma}\at_{\text{dec}} \theta^8 \,.
\ee
Moreover, the Hubble scale at the inflaton decay is given by:
\be
H_{\text{dec},\tau_2} = H_{\text{inf}}\, e^{-\frac{3}{2}N_{\tau_2}} \,,
\ee
allowing us to calculate the Hubble scale at radiation-volume equality:
\be
H_{\text{eq}} = H_{\text{dec},\tau_2}\, \theta^4 = 
H_{\rm inf}\, e^{-\frac{3}{2}N_{\tau_2}}\, \theta^4 \, .
\ee
Using the fact that the decay rate of the volume mode is:
\be
\Gamma_{\tau_b} \simeq c_{\text{loop}}^2 \left(\frac{m_{3/2}}{m_{\tau_b}}\right)^4 \frac{m_{\tau_b}^3}{M_p^2} \simeq c_{\text{loop}}^2 \frac{M_p}{\V^{5/2}} \,,
\ee
we can now estimate the number of efoldings of the matter epoch dominated by volume mode as:
\be
N_{\tau_b} = \ln \left(\frac{a_{\text{dec},\tau_b}}{a_{\text{eq}}}\right) \simeq \frac{2}{3}\ln\left(\frac{H_{\text{eq}}}{\Gamma_{\tau_b}}\right) \simeq \frac23 \ln\V - N_{\tau_2}\,,
\ee
where we considered $\theta^4 c_{\rm loop}^{-2}\sim\order{1}$. Therefore, the total number of efoldings of inflation becomes:
\be
N_e = 57 + \frac14 \ln r - \frac14 N_{\tau_2} -\frac14 N_{\tau_b} \simeq 50 - \frac14 N_{\tau_2}-\frac14 N_{\tau_b} \simeq 50 - \frac{1}{6}\ln\V\,.
\label{NeD7}
\ee
Note that this estimate gives a longer period of inflation with respect to the scenario where the inflaton is not wrapped by any D7 stack, even if there are two epochs of modulus domination. The reason is that both epochs, when summed together, last less that the single epoch of inflaton domination of the case with no D7-branes wrapped around the inflaton. As we shall see, this results in a better agreement with the observed value of the scalar spectral index. Lastly, we stress that the loop-enhanced volume mode coupling to the Higgs sector suppresses the production of axionic dark radiation. As stressed above, this coupling is however effective only when the Standard Model lives on D7-branes since it becomes negligible in sequestered scenarios where the visible sectors is localised on D3-branes at dP singularities. In this case the volume would decay into Higgs degrees of freedom via a Giudice-Masiero coupling \cite{Cicoli:2012aq, Allahverdi:2013noa, Cicoli:2022uqa} and a smaller decay width $\Gamma_{\tau_b}\sim M_p /\vo^{9/2}$ that would make the number of efoldings of inflation much shorter.
\end{itemize}

\subsection{Numerical examples}

\subsubsection{No D7s wrapped around the inflaton}

\begin{figure}[h]
    \centering
    \includegraphics[scale=0.8]{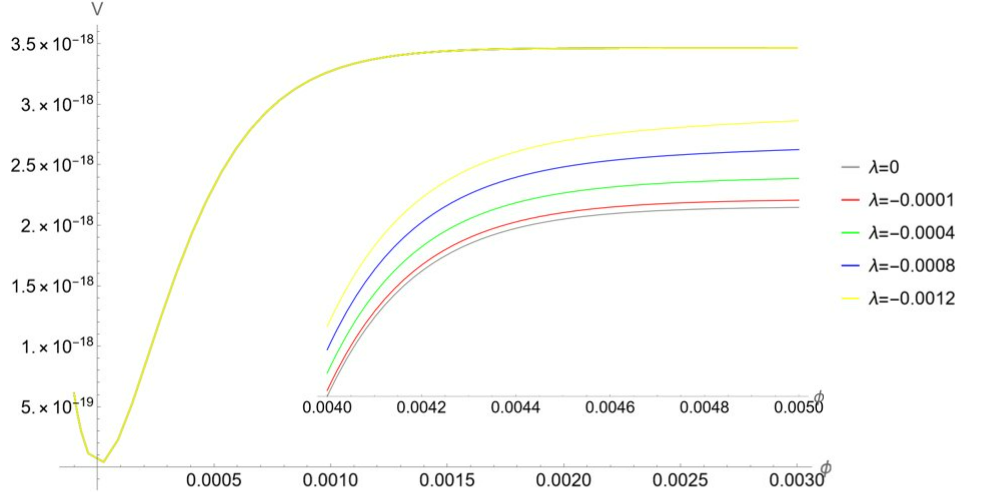}
\caption{Potential of blow-up inflation with $\Pi_2=-1$ and different values of $\lambda$. The difference between the corrections is visible in the zoomed region with $\phi \in [0.004;0.005]$}
    \label{BU_potential_plot}
\end{figure}

To quantitatively study the effect of higher derivative corrections, let us consider an explicit example characterised by the following choice of parameters:

\begin{center}
    \begin{tabular}{|c|c|c|c|c|c|c|c|c|}
    \hline
     $W_0$   &  $g_s$ & $\xi$ & $a_1$ & $a_2$ & $A_1$ & $A_2$ & $\beta_1$ & $\beta_2$\\
     \hline\hline
      0.1  &  0.13 & 0.1357 & $2\pi$ & $2\pi$ & 0.2 & $3.4\times 10^{-7}$ & 0.4725 & 0.01\\
      \hline
    \end{tabular}
\end{center}
For simplicity, we fix $\Pi_1=\Pi_b=0$ and the model is studied by varying $\Pi_2$ and $\lambda$. Let us stress that this assumption does not affect the main result since the leading $F^4$ correction is the one proportional to $\Pi_2$. Fig. \ref{BU_potential_plot} shows the plot of the uncorrected inflationary potential (gray line) which is compared with the corrected potential obtained by setting $\Pi_2=-1$ and choosing $\lambda \sim \mathcal{O}(10^{-4}-10^{-3})$.

\begin{figure}[h]
    \centering
    \includegraphics[scale=0.4]{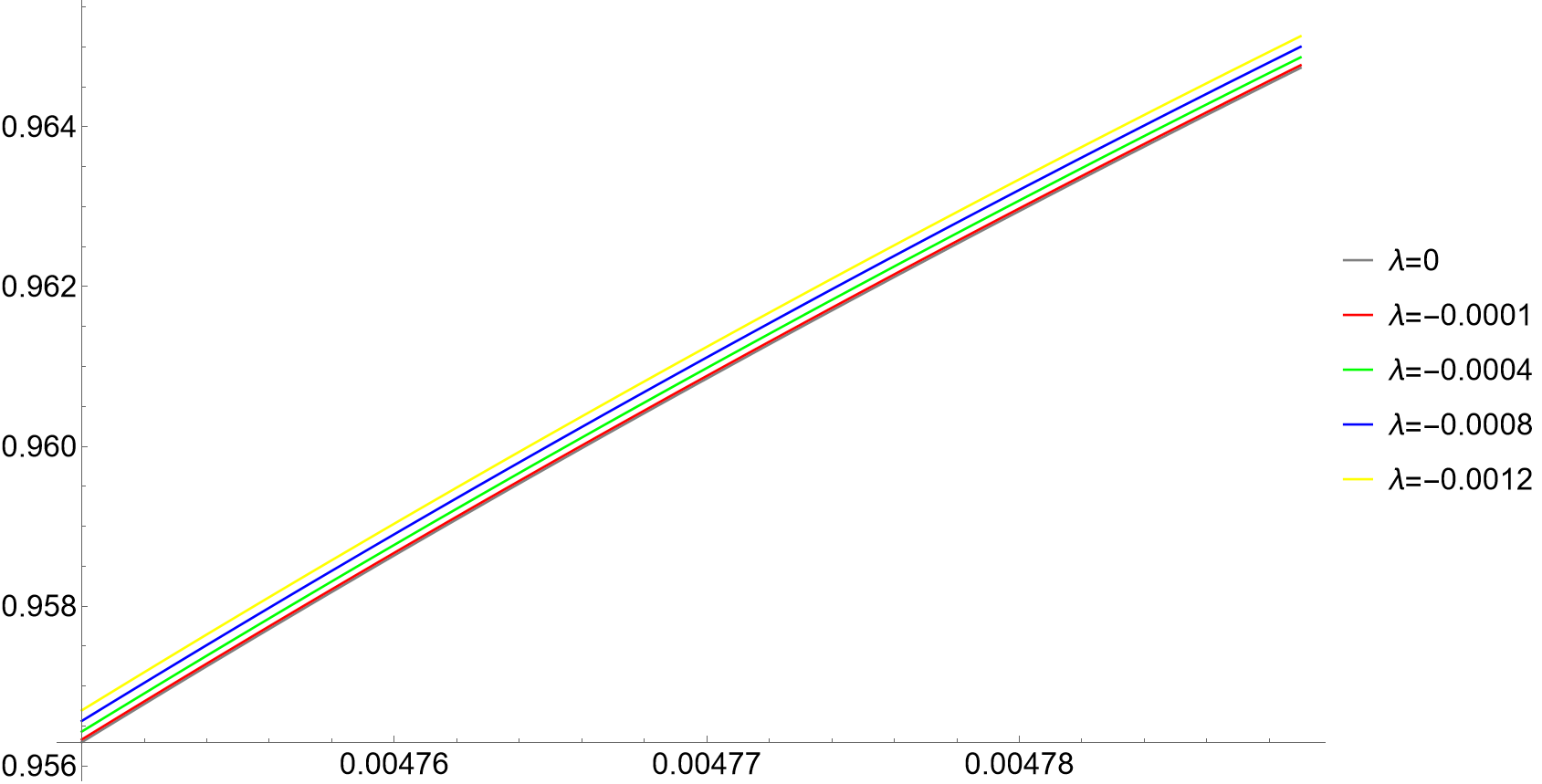}
\caption{Spectral index for different values of $\lambda$. The field at horizon exit is given in Tab. \ref{BU_tab}.}
    \label{BU_ns_plot}
\end{figure}

Knowing the explicit expression of the potential, we determine the spectral index (shown in Fig. \ref{BU_ns_plot} as function of $\phi$) and, by integration, the number of efoldings. In this scenario the inflaton is the longest-living particle and the number of efoldings to consider for inflation is $N_e=45.34$. Given the relations (\ref{NeNoD7}) and (\ref{NeEq}), we find the value of the field at horizon exit $\phi_{\rm exit}$, and then the value of the spectral index $n_s(\phi_{\rm exit})$ which is reported in Tab. \ref{BU_tab} for each value of $\lambda$.

In order for $n_s(\phi_{\rm exit})$ to be compatible with Planck measurements \cite{Planck:2018vyg}:
\begin{equation}
n_s = 0.9649 \pm 0.0042 \qquad  (68 \% \ \text{CL}) \,,
    \label{nsObs}
\end{equation}
we need to require $|\lambda| \lesssim 1.1 \times 10^{-3}$ for compatibility within $2\sigma$. This bound might be satisfied by actual multi-field models since, as can be seen from (\ref{eq:suppresion}), the single-field case features $|\lambda| = 3.5 \cdot 10^{-4}$ and, as already explained, we expect a similar suppression to persist also in the case with several moduli. 

\begin{table}[h!]
\centering
    \begin{tabular}{|c||c|c|c|}
    \hline
     $|\lambda|$   &  $\phi_{\rm exit}$ & $n_s$ & $A_s$ \\
     \hline\hline
      0  &  $4.494899\times10^{-3}$ & $0.956386$ & $2.11146\times10^{-9}$\\
      \hline
      $1.0\times10^{-4}$ &  $4.95668\times10^{-3}$ & $0.958164$ & $1.94664\times10^{-9}$\\
      \hline
      $4.0\times10^{-4}$ &  $4.98039\times10^{-3}$ & $0.963219$ & $1.53505\times10^{-9}$\\
      \hline
      $8.0\times10^{-4}$ &  $5.01349\times10^{-3}$ & $0.969316$ & $1.13485\times10^{-9}$\\
      \hline
      $1.2\times 10^{-3}$ &  $5.04829\times10^{-3}$ & $0.974691$ & $8.53024\times10^{-10}$\\
      \hline
    \end{tabular}
    \caption{Values of the inflaton at horizon exit $\phi_{\rm exit}$, the spectral index $n_s$ and the amplitude of the scalar perturbations $A_s$ for different choices of $\lambda$.\label{BU_tab}}
\end{table}

By comparing in Tab. \ref{BU_tab} the $\lambda=0$ case with the cases with non-zero $\lambda$, it is clear that $F^4$ corrections are a welcome effect, if $|\lambda|$ is not too large, since they can increase the spectral index improving the matching with CMB data. This is indeed the case when $\Pi_2$ is negative, as we have chosen. On the other hand, when $\Pi_2$ is positive, higher derivative $\alpha'^3$ corrections would induce negative corrections to $n_s$ that would make the comparison with actual data worse. Such analysis therefore suggests that geometries with negative $\Pi_2$ would be preferred in the context of blow-up inflation.

\subsubsection{D7s wrapped around the inflaton}

Let us now consider the scenario where the inflaton is wrapped by a stack of $N$ D7-branes supporting a gauge theory that undergoes gaugino condensation. As illustrative examples, we choose the following parameters:

\begin{center}
    \begin{tabular}{|c|c|c|c|c|c|c|c|c|}
    \hline
     $W_0$   &  $g_s$ & $\xi$ & $a_1$ & $a_2$ & $A_1$ & $A_2$ & $\beta_1$ & $\beta_2$\\
     \hline\hline
      0.1  &  0.13 & 0.1357 & $2\pi$ & $2\pi/N$ & 0.19 & $3.4\times 10^{-7}$ & $\simeq$ \ 0.5 & 0.01\\
      \hline
    \end{tabular}
\end{center}

Considering $N=2,3,5$, the total number of efoldings is now given by $N_e=47.90$ for $N=2$, $N_e=47.93$ for $N=3$, and $N_e=48.02$ for $N=5$. Repeating the same procedure as before for $\Pi_2 =-1$, we find the results shown in Tab.~\ref{BU_tab_N}. 

\begin{table}[h!]
    \centering
    \begin{tabular}{||c||c||c|c|c|}
    \hline
     $N$ & $|\lambda|$   &  $\phi_{\rm exit}$ & $n_s$ & $A_s$ \\
     \hline\hline
      \multirow{5}{3em}{$N=2$}& 0  &  $8.55743\times10^{-3}$ & $0.957692$ & $2.20009\times10^{-9}$\\
      
      & $1.0\times10^{-3}$ &  $8.59968\times10^{-3}$ & $0.962656$ & $1.73612\times10^{-9}$\\
     
      & $2.0\times10^{-3}$ &  $8.64364\times10^{-3}$ & $0.967233$ & $1.38248\times10^{-9}$\\
      
      & $3.0\times10^{-3}$ &  $8.68932\times10^{-3}$ & $0.971425$ & $1.11088\times10^{-9}$\\
      
      & $4.0\times 10^{-3}$ &  $8.73669\times10^{-3}$ & $0.975239$ & $9.00672\times10^{-10}$\\
      \hline
      \multirow{5}{3em}{$N=3$}& 0  &  $1.22049\times10^{-2}$ & $0.957679$ & $2.28554\times10^{-9}$\\
      
      & $2.0\times10^{-3}$ &  $1.22649\times10^{-2}$ & $0.96252$ & $1.81483\times10^{-9}$\\
     
      & $4.0\times10^{-3}$ &  $1.23273\times10^{-2}$ & $0.966995$ & $1.45345\times10^{-9}$\\
      
      & $6.0\times10^{-2}$ &  $1.23921\times10^{-2}$ & $0.971106$ & $1.174\times10^{-9}$\\
      
      & $8.0\times 10^{-3}$ &  $1.24593\times10^{-2}$ & $0.97486$ & $9.56344\times10^{-10}$\\
      \hline
      \multirow{5}{3em}{$N=5$}& 0  &  $1.78617\times10^{-2}$ & $0.957678$ & $2.14345\times10^{-9}$\\
      
      & $5.0\times10^{-3}$ &  $1.7953\times10^{-2}$ & $0.962446$ & $1.70842\times10^{-9}$\\
     
      & $1.0\times10^{-2}$ &  $1.80479\times10^{-2}$ & $0.966862$ & $1.37293\times10^{-9}$\\
      
      & $1.5\times10^{-2}$ &  $1.81464\times10^{-3}$ & $0.970927$ & $1.11241\times10^{-9}$\\
      
      & $2.0\times 10^{-2}$ &  $1.82484\times10^{-2}$ & $0.974647$ & $9.08706\times10^{-10}$\\
      \hline
    \end{tabular}
\caption{Values of the inflaton at horizon exit $\phi_{\rm exit}$, the spectral index $n_s$ and the amplitude of the scalar perturbations $A_s$ for different choices of $\lambda$ and $N=2,3,5$.\label{BU_tab_N}}
\end{table}

Due to a larger number of efoldings with respect to the case where the inflaton is not wrapped by any D7-stack, now the prediction for the spectral index falls within $2\sigma$ of the observed value also for $\lambda=0$. Non-zero values of $\lambda$ can improve the agreement with observations if $|\lambda|< |\lambda|_{\rm max}$ where: 

\begin{center}
    \begin{tabular}{|c||c|c|c|}
    \hline
&  $N=2$ & $N=3$ & $N=5$\\
\hline
$|\lambda|_{\rm max}$  &  $3.48 \times 10^{-3}$ & $7.15 \times 10^{-3}$ & $1.82 \times 10^{-2}$\\
      \hline 
    \end{tabular}
\end{center} 

In this case, given the larger number of efoldings, geometries with positive $\Pi_2$ can also be viable even if the corrections to the spectral index would be negative. Imposing again accordance with (\ref{nsObs}) at $2\sigma$ level for $\Pi_2=1$, we would obtain for example $|\lambda|_{\rm max}=2.29 \times 10^{-4}$ for $N=2$.

\section{Fibre inflation with $F^4$ corrections}
\label{sec_FI}

Similarly to blow-up inflation, the minimal version of fibre inflation \cite{Cicoli:2008gp,Cicoli:2016xae, Burgess:2016owb,Cicoli:2017axo,Cicoli:2018cgu,Bhattacharya:2020gnk,Cicoli:2020bao,Cicoli:2022uqa} involves also three K\"ahler moduli: two of them are stabilised via the standard LVS procedure and the remaining one can serve as an inflaton candidate in the presence of perturbative corrections to the K\"ahler potential. However, fiber inflation requires a different geometry from the one of blow-up inflation since one needs CY threefolds which are K3 fibrations over a $\mathbb{P}^1$ base. The simplest model requires the addition of a blow-up mode such that the volume can be expressed as:
\begin{equation}
\mathcal{V}=\frac{1}{6}\left(k_{111}(t^1)^3+3k_{233}t^2(t^3)^2\right)=\alpha\left(\sqrt{\tau_2}\tau_3-\tau_1^{3/2}\right).
\end{equation}
The requirement of having a K3 fibred CY threefold with at least a ddP$_n$ divisor for LVS moduli stabilisation is quite restrictive. The corresponding scanning results for the number of CY geometries suitable for realising fibre inflation are presented in Tab.~\ref{tab_fibre-Gstar}.

\begin{table}[H]
 \centering
\hskip0.11cm \begin{tabular}{|c|c|c||c|c|c||c|}
\hline
 $h^{1,1}$ & Poly$^*$ &  Geom$^*$ & $n_{\rm LVS}$ & K3 fibred  & $n_{\rm LVS}$ with K3 fib.& $n_{\rm LVS}$ with \\
   &  & $(n_{\rm CY})$  & &  CY & (fibre inflation)  &  K3 fib. \& $D_\Pi$ \\
 \hline
 1 & 5 & 5 & 0 & 0  & 0  &  0  \\
 2 & 36 & 39 & 22 & 10  & 0  &  0  \\
 3 & 243 & 305 & 132 & 136  & 43 &  0  \\
 4 & 1185 & 2000 & 750 & 865 & 171  & 28  \\
 5 & 4897 & 13494 & 4104  & 5970  &  951 & 179 \\
 \hline
  \end{tabular}
  \caption{Number of LVS CY geometries suitable for fibre inflation.}
    \label{tab_fibre-Gstar}
 \end{table}

It is worth mentioning that the scanning results presented in Tab. \ref{tab_fibre-Gstar} are consistent with the previous scans performed in \cite{Cicoli:2016xae, Cicoli:2011it}. To be more specific, the number of distinct K3 fibred CY geometries supporting LVS was found in \cite{Cicoli:2016xae} to be 43 for $h^{1,1}=3$, and ref. \cite{Cicoli:2011it} claimed that the number of polytopes giving K3 fibred CY threefolds with $h^{1,1}=4$ and at least one diagonal del Pezzo ddP$_n$ divisor is 158.

\subsection{Inflationary potential}

The leading order scalar potential of fibre inflation turns out to be:
\begin{equation}
V(\mathcal{V},\tau_1)=a^2_1 |A_1|^2\frac{\sqrt{\tau_1}}{\mathcal{V}}e^{-2a_1\tau_1}-a_1\, |A_1|\,|W_0|\frac{\tau_1}{\mathcal{V}}e^{-a_1\tau_1}+\frac{\xi \,|W_0|^2}{g_s^{3/2}\mathcal{V}^3}\,,
 \end{equation}
with a flat direction in the $(\tau_2,\tau_3)$ plane which plays the role of the inflaton (the proper canonically normalised inflationary direction orthogonal to the volume mode is given by the ratio between $\tau_2$ and $\tau_3$). The inflaton potential is generated by subdominant string loop corrections:
\begin{equation}
\label{FI_tau}
    \delta V_{\mathcal{O}(\mathcal{V}^{-10/3})}(\tau_2)=\left(g_s^2\frac{A}{\tau_1^2}-\frac{B}{\mathcal{V}\sqrt{\tau_2}}+g_s^2\frac{C\tau_2}{\mathcal{V}^2}\right)\frac{|W_0|^2}{\mathcal{V}^2}\,,
\end{equation}
where $A, B, C$ are flux-dependent coefficients that are expected to be of $\mathcal{O}(1)$. The minimum of this potential is approximately located at:
\begin{equation}
\braket{\tau_2} \simeq g_s^{4/3}\left(\frac{4A}{B}\right)^{2/3}\braket{\mathcal{V}}^{2/3} \,.
\end{equation}
Writing the canonically normalised inflaton field $\phi$ as:
\begin{equation}
\tau_2=\braket{\tau_2}e^{\frac{2\hat{\phi}}{\sqrt{3}}} \simeq  g_s^{4/3}\left(\frac{4A}{B}\right)^{2/3}\braket{\mathcal{V}}^{2/3} e^{\frac{2\hat{\phi}}{\sqrt{3}}}\,,
\label{tau2inf}
\end{equation}
where $\hat{\phi}$ is the shift with respect to the minimum, i.e. $\phi=\braket{\phi}+\hat{\phi}$,  the potential (\ref{FI_tau}) becomes:
\begin{equation}
\label{FI_phi}
    V_{\rm inf}(\hat{\phi})=V_0\left[3-4e^{-\hat{\phi}/\sqrt{3}}+e^{-4\hat{\phi}/\sqrt{3}}+R\left(e^{2\hat{\phi}/\sqrt{3}}-1\right)\right],
\end{equation}
where (introducing a proper normalisation factor $g_s/(8\pi)$ from dimensional reduction):
\begin{equation}
\label{V0_R}
V_0\equiv \frac{g_s^{1/3}\,|W_0|^2A}{8\pi\braket{\mathcal{V}}^{10/3}}\left(\frac{B}{4A}\right)^{4/3} \qquad\text{and}\qquad  R \equiv 16g_s^4 \frac{AC}{B^2}\ll 1 \,.
\end{equation}
Note that we added in (\ref{FI_phi}) an uplifting term to obtain a Minkowski vacuum. The slow-roll parameters derived from the inflationary potential look like:
\begin{eqnarray}
\epsilon(\hat{\phi})&=&\frac{2}{3}\frac{\left(2e^{-\hat{\phi}/\sqrt{3}}-2e^{-4\hat{\phi}/\sqrt{3}}+Re^{2\hat{\phi}/\sqrt{3}}\right)^2}{\left(3-R+e^{-4\hat{\phi}/\sqrt{3}}-4e^{-\hat{\phi}/\sqrt{3}}+Re^{2\hat{\phi}/\sqrt{3}}\right)^2}\,, \\
\eta(\hat{\phi})&=&\frac{4}{3}\frac{4e^{-4\hat{\phi}/\sqrt{3}}-e^{-\hat{\phi}/\sqrt{3}}+Re^{2\hat{\phi}/\sqrt{3}}}{\left(3-R+e^{-4\hat{\phi}/\sqrt{3}}-4e^{-\hat{\phi}/\sqrt{3}}+Re^{2\hat{\phi}/\sqrt{3}}\right)}\,,
\end{eqnarray}
and the number of efoldings is:
\begin{equation}
N_e(\hat{\phi}_{\rm exit})=\int_{\hat{\phi}_{\rm end}}^{\hat{\phi}_{\rm exit}} \frac{1}{\sqrt{2\epsilon(\hat{\phi}})} \simeq \int_{\hat{\phi}_{\rm end}}^{\hat{\phi}_{\rm exit}} \frac{\left(3-4e^{-\hat{\phi}/\sqrt{3}}+Re^{2\hat{\phi}/\sqrt{3}}\right)}{\left(2e^{-\hat{\phi}/\sqrt{3}}+Re^{2\hat{\phi}/\sqrt{3}}\right)}\,,
\end{equation}
where $\hat{\phi}_{\rm end}$ and $\hat{\phi}_{\rm exit}$ are respectively the values of the inflaton at the end of inflation and at horizon exit.

\subsection{$F^4$ corrections}

Explicit CY examples of fibre inflation with chiral matter have been presented in \cite{Cicoli:2017axo} that has already stressed the importance to control $F^4$ corrections to the inflationary potential since they could spoil its flatness. This is in particular true for K3 fibred CY geometries since $\Pi({\rm K3}) = 24$, and so the coefficient of $F^4$ effects is non-zero. On the other hand, the theorem of \cite{Oguiso1993, Schulz:2004tt} allows in principle also for CY threefolds that are ${\mathbb T}^4$ fibrations over a $\mathbb{P}^1$ base. This case would be more promising to tame $F^4$ corrections since their coefficient would vanish due to $\Pi({\mathbb T}^4)=0$. However, in our scan for CY threefolds in the KS database we did not find any example with a ${\mathbb T}^4$ divisor. Thus, in what follows we shall perform a numerical analysis of fibre inflation with non-zero $F^4$ terms to study in detail the effect of these corrections on the inflationary dynamics.

\subsubsection*{Case 1: a single K3 fibre}

The minimal fibre inflation case is a three field model based on a CY threefold that features a K3-fibration structure with a diagonal del Pezzo divisor. Considering an appropriate basis of divisors, the intersection polynomial can be brought to the following form:
\be
I_3 = k_{111} \, D_{1}^3 + k_{233} \, D_{2}\,D_3^2 \,.
\ee
As the $D_2$ divisor appears linearly, from the theorem of \cite{Oguiso1993, Schulz:2004tt}, this CY threefold is guaranteed to be a K3 or ${\mathbb T}^4$ fibration over a ${\mathbb P}^1$ base. Furthermore, the triple-intersection number $k_{111}$ is related to the degree of the del Pezzo divisor $D_1=dP_{n}$ as $k_{111} = 9 - n$, while $k_{233}$ counts the intersections of the K3 surface $D_2$ with $D_3$. This leads to the following volume form:
\be
{\cal V} = \frac{k_{111}}{6} \, (t^1)^3 + \frac{k_{233}}{2} \, t^2 \,  (t^{3})^2 = \beta_2\, \, \sqrt{\tau_{2}}\,\tau_{3} - \beta_1 \, \, \tau_{1}^{3/2}  \, ,
\ee
where $\beta_1 = \frac{1}{3} \sqrt{\frac{2}{k_{111}}}$ and $\beta_2 = \frac{1}{\sqrt{2\, k_{233}}}$, and the 2-cycle moduli $t^i$ are related to the 4-cycle moduli $\tau_i$ as follows:
\be
t^1 = - \, \sqrt{\frac{2\,\tau_{1}}{k_{111}}} \,, \qquad t^2 = \, \frac{\tau_3}{\sqrt{2\,k_{233}\,\tau_2}}, \qquad t^3 = \sqrt{\frac{2\,\tau_{2}}{k_{233}}}\,.
\ee
The higher derivative $\alpha'^3$ corrections can be written as:
\bea
V_{F^4} &=& \frac{\gamma}{{\cal V}^4} \, \left( \Pi_1 \, t^1 + \, \Pi_2 \, t^2 + \Pi_3\, t^3 \right) \nonumber\\
&=&  \frac{\gamma}{{\cal V}^4} \, \left[\Pi_3\,  \sqrt{\frac{2\,\tau_{2}}{k_{233}}}  + \, \Pi_2 \, \left(\frac{\cal V}{\tau_2} + \frac{1}{3}\, \sqrt{\frac{2}{k_{111}}} \, \frac{\tau_1^{3/2}}{\tau_2} \right) - \Pi_1 \, \sqrt{\frac{2\,\tau_{1}}{k_{111}}} \right].
\label{VF4FI}
\eea
In the inflationary regime, $\vo$ is kept constant at its minimum while $\tau_2$ is at large values away from its minimum, as can be seen from (\ref{tau2inf}) for $\hat\phi>0$. Thus, the leading order term in (\ref{VF4FI}) is the one proportional to $\Pi_3$. Therefore, a leading order protection of the fibre inflation model can be guaranteed by demanding a geometry with $\Pi_3 = 0$. However, the subleading contribution proportional to $\Pi_2$ would still induce an inflaton-dependent correction that might be dangerous. The ideal situation to completely remove higher derivative $F^4$ corrections to fibre inflaton is therefore characterised by:
\be
\Pi_2 = \Pi_3 = 0\,,
\ee
where, as pointed out above, $\Pi_2$ would vanish for ${\mathbb T}^4$ fibred CY threefolds. Interestingly, such CY examples with ${\mathbb T}^4$ divisors have been found in the CICY database, without however any ddP for LVS \cite{Carta:2022web}. It is also true that all K3 fibred CY threefolds do not satisfy $\Pi_2 = 0$.

\subsubsection*{Case 2: multiple K3 fibres}

More generically, fibre inflation could be realised also in CY threefolds which admit multiple K3 or ${\mathbb T}^4$ fibrations together with at least a diagonal del Pezzo divisor. The corresponding intersection polynomial would look like (see \cite{Cicoli:2017axo} for explicit CY examples):
\be
I_3 = k_{111} \, D_{1}^3 + k_{234} \, D_{2}\,D_3\, D_4 \,.
\ee
As the divisors $D_2$, $D_3$ and $D_4$ all appear linearly, from the theorem of \cite{Oguiso1993, Schulz:2004tt}, this CY threefold is guaranteed to have three K3 or ${\mathbb T}^4$ fibrations over a ${\mathbb P}^1$ base. As before, $D_1$ is a  diagonal dP$_n$ divisor with $k_{111} = 9 - n > 0$. The volume form becomes:
\be
{\cal V} = \frac{k_{111}}{6} \, (t^1)^3 + k_{234}\, t^2 \, t^{3} \, t^4 = \beta_2\, \, \sqrt{\tau_{2}\,\tau_{3}\,\tau_{4}} - \beta_1 \, \, \tau_{1}^{3/2}  \, ,
\ee
where $\beta_1 = \frac{1}{3} \sqrt{\frac{2}{k_{111}}}$ and $\beta_2 = \frac{1}{\sqrt{k_{234}}}$, and the 2-cycle moduli $t^i$ are related to the 4-cycle moduli $\tau_i$ as:
\be
t^1 = - \, \sqrt{\frac{2\,\tau_{1}}{k_{111}}} \,, \qquad t^2 = \, \frac{\sqrt{\tau_3 \, \tau_4}}{\sqrt{k_{234}\,\tau_2}}, \qquad t^3 = \, \frac{\sqrt{\tau_2 \, \tau_4}}{\sqrt{k_{234}\,\tau_3}}, \qquad t^4 = \, \frac{\sqrt{\tau_2 \, \tau_3}}{\sqrt{k_{234}\,\tau_4}}\,.
\ee
This case features two flat directions which can be parametrised by $\tau_2$ and $\tau_2$. Moreover, the higher derivative $F^4$ corrections take the form:
\bea
\label{VF4-fibre2}
V_{F^4} &= & \frac{\gamma}{{\cal V}^4} \, \left( \Pi_1 \, t^1 + \, \Pi_2 \, t^2 + \Pi_3\, t^3 + \Pi_4\, t^4 \right) \nonumber\\
& = & \frac{\gamma}{{\cal V}^4} \, ({\cal V}+\beta_1 \, \, \tau_{1}^{3/2}) \left(\frac{\Pi_2}{\tau_2} +\, \frac{\Pi_3}{\tau_3} +\, \frac{\Pi_4\, \tau_2 \, \tau_3}{\beta_2\, ({\cal V}+\beta_1 \, \, \tau_{1}^{3/2})^2}\right) - \Pi_1 \,\frac{\gamma}{{\cal V}^4} \,  \sqrt{\frac{2\,\tau_{1}}{k_{111}}}\,.
\eea
In the explicit model of \cite{Cicoli:2017axo}, non-zero gauge fluxes generate chiral matter and a moduli-dependent Fayet-Iliopoulos term which lifts one flat direction, stabilising $\tau_3 \propto \tau_2$. After performing this substitution in (\ref{VF4-fibre2}), this potential scales as the one in the single field case given by (\ref{VF4FI}). Interestingly, ref. \cite{Cicoli:2017axo} noticed that, in the absence of winding string loop corrections, $F^4$ effects can also help to generate a post-inflationary minimum. Note finally that if all the divisors corresponding to the CY multi-fibre structure are ${\mathbb T}^4$, the $F^4$ terms would be absent. However, incorporating a diagonal del Pezzo within a ${\mathbb T}^4$-fibred CY is yet to be constructed (e.g. see \cite{Carta:2022web}).

\subsection{Constraints on inflation}

Let us focus on the simplest realisation of fibre inflation, and add the dominant $F^4$ corrections (\ref{VF4FI}) to the leading inflationary potential (\ref{FI_phi}). The total inflaton-dependent potential takes therefore the form:
\begin{equation}
\label{FIcorr_phi}
V_{\rm inf}(\hat{\phi})=V_0\left[e^{-4\hat{\phi}/\sqrt{3}}-4e^{-\hat{\phi}/\sqrt{3}}+3+R(e^{2\hat{\phi}/\sqrt{3}}-1)-R_2\,e^{-2\hat{\phi}/\sqrt{3}}-R_3\,e^{\hat{\phi}/\sqrt{3}}\right],
\end{equation}
where $R$ is given by (\ref{V0_R}) while $R_2$ and $R_3$ are defined as:
\begin{equation}
R_2\equiv  \frac{|W_0|^2 }{(4\pi)^2Ag_s^{3/2}}\frac{\lambda\Pi_2}{\mathcal{V}}\ll 1 \qquad\text{and}\qquad R_3 \equiv \frac{4\,|W_0|^2 \sqrt{g_s}}{B} \frac{\lambda\Pi_3}{\mathcal{V}}\ll 1 \,.
\end{equation}
Note that the most dangerous term that could potentially spoil the flatness of the inflationary plateau is the one proportional to $R_3$ since it multiplies a positive exponential. The term proportional to $R_2$ is instead harmless since it multiplies a negative exponential. 

As we have seen for blow-up inflation, the study of reheating after the end of inflation is crucial to determine the number of efoldings of inflation which are needed to make robust predictions for the main cosmological observables. Reheating for fibre inflation with the Standard Model on D7-branes has been studied in \cite{Cicoli:2018cgu}, while ref. \cite{Cicoli:2022uqa} analysed the case where the visible sector is realised on D3-branes. In both cases, a radiation dominated universe is realised from the perturbative decay of the inflaton after the end of inflation. In what follows we shall focus on the D7-brane case and include the loop-induced coupling between the inflaton and the Standard Model Higgs, similarly to volume-Higgs coupling found in \cite{Cicoli:2022fzy}. The relevant term in the low-energy Lagrangian is the Higgs mass term which can be expanded as:
\begin{equation}
m^2_h h^2 = m^2_{3/2}\left[c_0 +c_{\rm loop} \ln{\left(\frac{M_{KK}}{m_{3/2}}\right)} \right] h^2\,,
\label{Higgsmass}
\end{equation}
where $\ln\left(M_{KK}/m_{3/2}\right)\propto \ln{\mathcal{V}}$. Using the fact that \cite{Cicoli:2012cy}:
\begin{equation}
\mathcal{V}= \braket{\mathcal{V}}(1+ \kappa \hat{\phi})\,,
\end{equation}
with $\kappa \sim \langle\mathcal{V}\rangle^{-1/3}$, the Higgs mass term (\ref{Higgsmass}) generates a coupling between $\hat\phi$ and $h$ that leads to the following decay rate:
\begin{equation}
\Gamma_{\phi\rightarrow hh} \simeq \frac{c_{\rm loop}^2}{\vo^{2/3}}\frac{m_{3/2}^4}{M_p^2 m_{\rm inf}} \simeq \frac{c_{\rm loop}^2}{\vo^{2/3}} \left(\frac{m_{3/2}}{m_{\rm inf}} \right)^4 \Gamma_{\phi \rightarrow \gamma\gamma} \,.
\label{GammaHiggs}
\end{equation}
It is then easy to realise that the inflaton decay width into Higgses is larger than the one into gauge bosons for $\vo\gg 1$ since:
\begin{equation}
\frac{\Gamma_{\phi\rightarrow hh}}{\Gamma_{\phi \rightarrow \gamma\gamma}} \simeq (c_{\rm loop} \mathcal{V})^2 \gg 1\,.
\label{GammaRatio}
\end{equation} 
The number of efoldings of inflation is then determined as:
\begin{equation}
N_e \simeq 56 - \frac{1}{3}\ln{ \left(\frac{m_{\rm inf}}{T_{\rm rh}}\right)}\,,
\label{NeFI}
\end{equation}
where the reheating temperature $T_{\rm rh}$ scales as:
\be
T_{\rm rh} \simeq \sqrt{\Gamma_{\phi\rightarrow hh}\,M_p}\,.
\ee
Substituting this expression in (\ref{NeFI}), and using (\ref{GammaHiggs}), we finally find:
\be
N_e \simeq 53 + \frac16 \ln\left[1 + \frac{c^2_{\rm loop}}{\mathcal{V}^{2/3}}\left(\frac{m_{3/2}}{m_{\rm inf}}\right)^4\right]. 
\label{NeFinalFI}
\ee
This is the number of efoldings of inflation used in the next section for the analysis of the inflationary dynamics in some illustrative numerical examples.

\subsection{Numerical examples}

Let us now perform a quantitative study of the effect of higher derivative $\alpha'^3$ corrections to fibre inflation for reasonable choices of the underlying parameters. In order to match observations, we follow the best-fit analysis of \cite{Cicoli:2020bao} and set $R= 4.8\times 10^{-6}$, which can be obtained by choosing:
\begin{equation}
A=1\,, \qquad B=8\,, \qquad C=0.19 \,.
\end{equation}
Moreover, given that $D_2$ is a K3 divisor, we fix $\Pi_2 = 24$, while we leave $\Pi_3$ and $\lambda$ as free parameters that we constrain from phenomenological data.

\begin{figure}[h]
    \centering
    \includegraphics[scale=0.4]{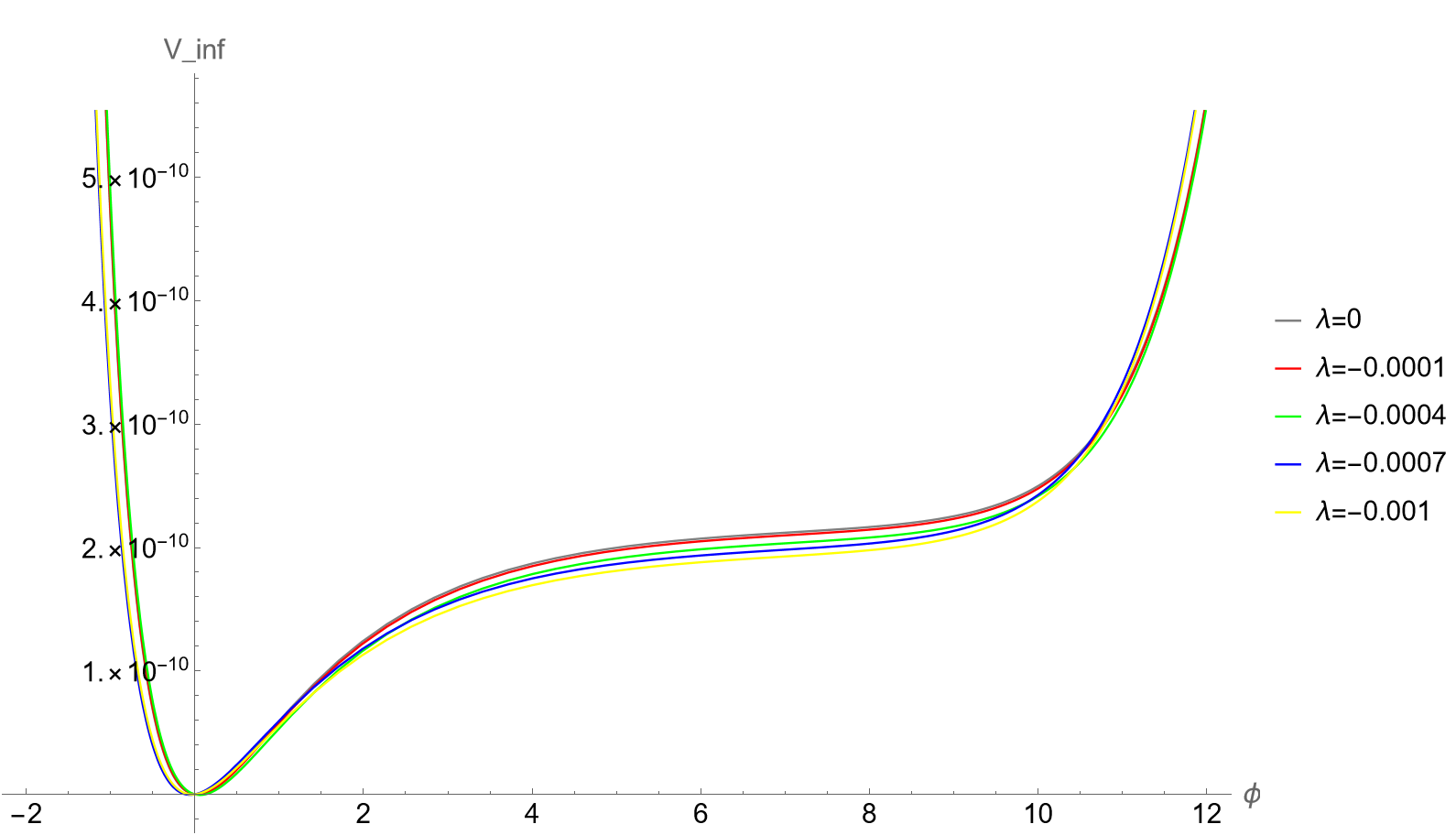}
\caption{Potential of fibre inflation with $F^4$ corrections with $\Pi_3=1$ and different values of $\lambda$.}
    \label{FI_potential_plot}
\end{figure}

Fig. \ref{FI_potential_plot} shows the potential of fibre inflation with $F^4$ corrections corresponding to $\Pi_3=1$ and different negative values of $\lambda$. 

As for blow-up inflation, we find numerically the range of values of $\lambda$ which are compatible with observations. In Tab. \ref{FI_tab} we show the values for the spectral index evaluated at horizon exit, with $N_e=53.81$ fixed from (\ref{NeFinalFI}), for $\Pi_3=1$ and different values of $\lambda$. In order to reproduce the best-fit value of the scalar spectral index \cite{Cicoli:2020bao, Planck:2018vyg}:
\be
n_s = 0.9696^{+0.0010}_{-0.0026}
\ee
the numerical coefficient $\lambda$ has to respect the bound $|\lambda|\lesssim  6.1 \times 10^{-4}$, which seems again compatible with the single-field result (\ref{eq:suppresion}) that gives $|\lambda| = 3.5 \cdot 10^{-4}$.

\begin{figure}[h!]
\centering
\includegraphics[scale=0.4]{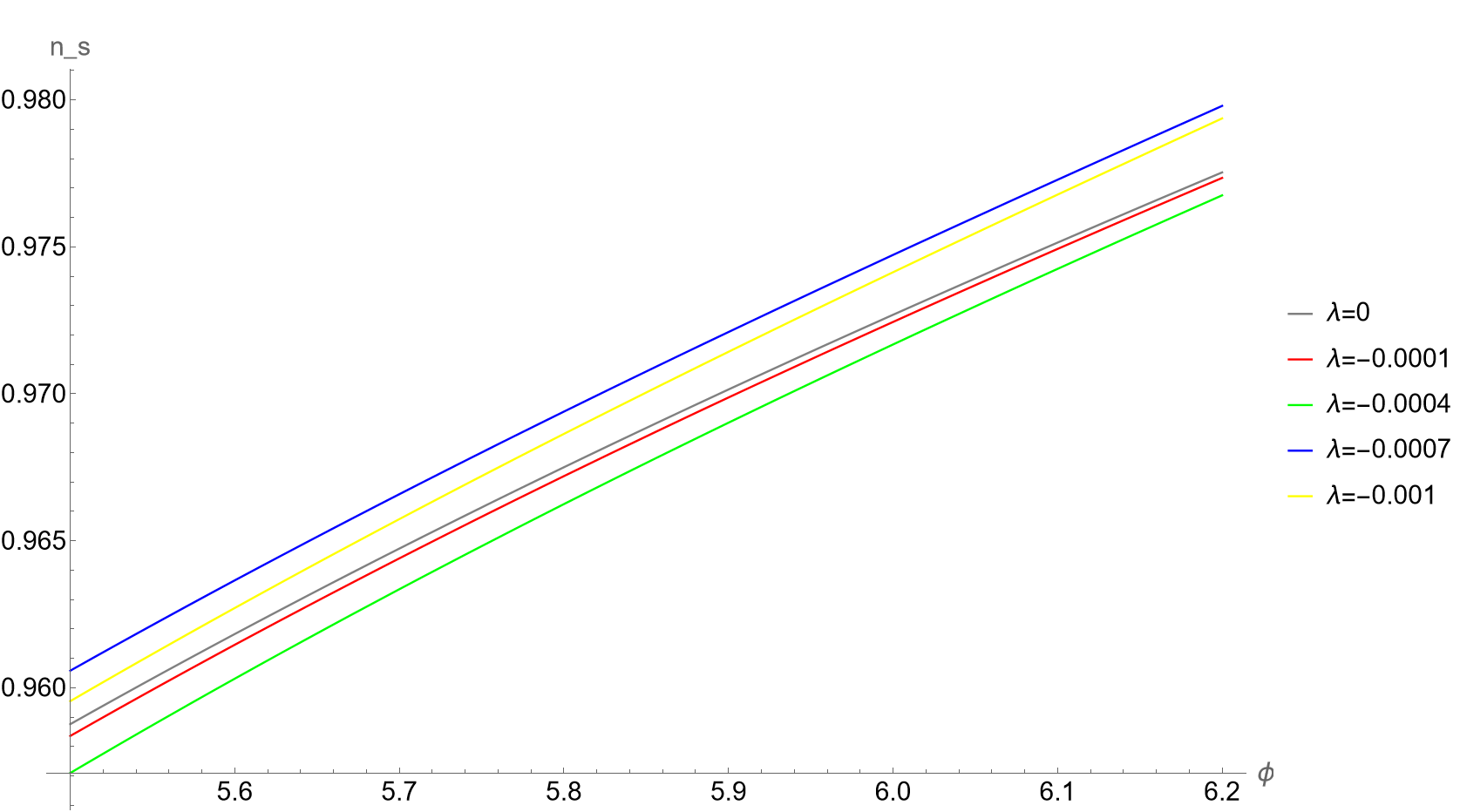}
\caption{Spectral index for different values of $\lambda$ in fibre inflation. The value of the inflaton at horizon exit is given in Tab. \ref{FI_tab}.}
\label{FI_ns_plot}
\end{figure}

\begin{table}[h!]
\centering
\begin{tabular}{|c||c|c|c|}
\hline
$|\lambda|$   &  $\phi_{\rm exit}$ & $n_s$ & $A_s$ \\
\hline\hline
0  &  $5.91328$ & $0.97049$ & $2.13082\times10^{-9}$\\
\hline
$0.1\times10^{-3}$ &  $5.93005$ & $0.970657$ & $2.09702\times10^{-9}$\\
\hline
$0.4\times10^{-3}$ &  $5.98203$ & $0.971207$ & $1.99576\times10^{-9}$\\
\hline
$0.7\times10^{-3}$ &  $5.88793$ & $0.97178$ & $1.90293\times10^{-9}$\\
\hline
$1.0\times 10^{-3}$ &  $5.93552$ & $0.972399$ & $1.81416\times10^{-9}$\\
\hline
\end{tabular}
\caption{Values of the inflaton at horizon exit $\phi_{\rm exit}$, the spectral index $n_s$ and the amplitude of the scalar perturbations $A_s$ for different choices of $\lambda$ in fibre inflation.\label{FI_tab}}
\end{table}

\section{Poly-instanton inflation with $F^4$ corrections}
\label{sec_PI}

Let us finally analyse higher derivative $\alpha'^3$ corrections to poly-instanton inflation, focusing on its simplest realisation based on a three-field LVS model \cite{Cicoli:2011ct, Blumenhagen:2012ue}. This model involves exponentially suppressed corrections appearing on top of the usual non-perturbative superpotential effects arising from E3-instantons or gaugino condensation wrapping suitable rigid cycles of the CY threefold. In this three-field model, two K\"ahler moduli correspond to the volumes of the `big' and `small' 4-cycles (namely $D_b$ and $D_s$) of a typical Swiss-cheese CY threefold, while the third modulus controls the volume of a Wilson divisor $D_w$ which is a  ${\mathbb P}^1$ fibration over ${\mathbb T}^2$ \cite{Blumenhagen:2012kz}. Moreover, such a divisor has the following Hodge numbers for a specific choice of involution: $h^{2,0}(D_w) = 0$ and $h^{0,0}(D_w)= h^{1,0} (D_w)= h^{1,0}_+(D_w)=1$. For this model one can consider the following intersection polynomial:
\be
I_3 = k_{sss} \, D_{s}^3 + k_{ssw} \, D_s^2\, D_{w} + k_{sww} \, D_s\, D_w^2 +  k_{bbb} \, D_{b}^3  \,, 
\ee
where, as argued earlier, the self triple-intersection number of the Wilson divisor is zero, i.e. $k_{www} =0$. This is because Wilson divisors are of the kind given in (\ref{eq:D3PibothZero}) for $n=0$. We also have selected a basis of divisors where the large four-cycle $D_b$ does not mix with the other two divisors to keep a strong Swiss-cheese structure. This leads to the following form of the CY volume:
\be
{\cal V} = \frac{k_{bbb}}{6} \, (t^b)^3 + \frac{k_{sss}}{6} \, (t^s)^3 + \frac{k_{ssw}}{2} \, (t^s)^2\, t^w  + \frac{k_{sww}}{2} \, t^s\, (t^w)^2 \,,
\ee
which subsequently gives to the following 4-cycle volumes:
\bea
\tau_b &=& \frac{1}{2}\, k_{bbb}\, (t^b)^2, \quad \tau_s = \frac{1}{2}\, k_{sss} \left((t^s)^2 + 2\, \frac{k_{ssw}}{k_{sss}}\, (t^s)^2 + \frac{k_{sww}}{k_{sss}}\, (t^w)^2 \right), \nonumber\\
\tau_w &=& \frac{1}{2}\, \left(k_{ssw}\, (t^s)^2 + 2\, k_{sww}\, t^s\, t^w\right)\,.
\eea
Now it is clear that in order for the `small' divisor to be diagonal, the above intersection numbers have to satisfy the following relation:
\be
k_{sss} = \pm\, k_{ssw} = k_{sww} \,,
\ee
which is indeed the case when the divisor basis is appropriately chosen in the way we have described above. This leads to the following expression of the CY volume:
\be
{\cal V} =  \beta_b \, \, \tau_{b}^{3/2} - \beta_s \, \, \tau_{s}^{3/2} - \beta_s\, \, (\tau_s \mp \tau_w)^{3/2}\,,
\ee
where $\beta_s = \frac{1}{3} \sqrt{\frac{2}{k_{sss}}}$ and $\beta_b = \frac{1}{3} \sqrt{\frac{2}{k_{bbb}}}$, and the 4-cycle volumes $\tau_s$, $\tau_w$ and $\tau_b$ are given by:
\be\tau_b = \frac{1}{2}\, k_{bbb}\, (t^b)^2, \quad \tau_s = \frac{1}{2}\, k_{sss} \, (t^s \pm t^w)^2, \quad \tau_w = \pm \frac{1}{2}\, k_{sss} \, (t^s \pm 2\, t^w)\,.
\ee
The $\pm$ sign is decided by the K\"ahler cone conditions, like for example in the case of $D_s$ being a del Pezzo divisor where the corresponding two-cycle in the K\"ahler form $J$ satisfies $t^s <  0$ in an appropriate diagonal basis. Looking at explicit CY examples \cite{Blumenhagen:2012kz}, the sign is fixed through the K\"ahler cone conditions such that $k_{sss} = - \, k_{ssw} = k_{sww}$, leading to the following peculiar structure of the volume form \cite{Blumenhagen:2012kz}:
\bea
{\cal V} &=&  \beta_b \, \, \tau_{b}^{3/2} - \beta_s \, \, \tau_{s}^{3/2} - \beta_s\, \, (\tau_s + \tau_w)^{3/2}\,, \\
\tau_b &=& \frac{1}{2}\, k_{bbb}\, (t^b)^2, \quad \tau_s = \frac{1}{2}\, k_{sss} \, (t^s - t^w)^2, \quad \tau_w = - \, \frac{1}{2}\, k_{sss} \, (t^s -\, 2\, t^w)\,. \nonumber
\eea

\subsection{Divisor topologies for poly-instanton inflation}

In principle, one should be able to fit the requirements for poly-instanton inflation on top of having LVS in a setup with three K\"ahler moduli. Indeed we find that there are four CY threefold geometries with $h^{1,1}(X) = 3$ in the KS database which have exactly one Wilson divisors and a ${\mathbb P}^2$ divisor. However, as mentioned in \cite{Blumenhagen:2012kz}, in order to avoid all vector-like zero modes to have poly-instanton effects, one should ensure that the rigid divisors wrapped by the ED3-instantons, should have some orientifold-odd $(1,1)$-cycles which are trivial in the CY threefold. Given that ${\mathbb P}^2$ has a single $(1,1)$-cycle, it would certainly not have such additional two-cycles which could be orientifold-odd and then trivial in the CY threefold. Hence one has to look for CY examples with $h^{1,1}(X) = 4$ for a viable model of poly-instanton inflation as presented in \cite{Blumenhagen:2012kz,Blumenhagen:2012ue}. In this regard, we present the classification of all CY geometries relevant for LVS poly-instanton inflation in Tab. \ref{tab_GwilsonLVS}. 

Let us stress that in all our scans we have only focused on the minimal requirements to realise explicit global constructions of LVS inflationary models. However, every model has to be engineered in a specific way on top of fulfilling the first order topological requirements, as we do. For example, merely having a K3-fibred CY threefold with a diagonal del Pezzo for LVS does not guarantee a viable fibre inflation model until one ensures that string loop corrections can appropriately generate the right form of the scalar potential after choosing some concrete brane setups.

\begin{table}[H]
  \centering
 \begin{tabular}{|c||c|c||c|c|c||c|c|c|c|}
\hline
$h^{1,1}$ & Poly$^*$  & Geom$^*$  & Single & Two & Three  & $n_{\rm LVS}$ & $n_{\rm LVS}$ \& $W$ & $n_{\rm LVS}$ \&  $W_\Pi$   \\
&  & $(n_{\rm CY})$ & $W$ & $W$ & $W$  &  & (poly-inst.) & (topol. tamed)     \\
 \hline
 1 & 5 & 5 & 0 & 0  & 0 & 0 & 0  & 0    \\
 2 & 36 & 39 & 0 & 0 & 0 & 22 & 0  & 0  \\
 3 & 243 & 305 & 19 & 0 & 0 & 132 & 4 & 4   \\
 4 & 1185 & 2000 & 221 & 8 & 0 & 750 & 75  & 63   \\
 5 & 4897 &13494 & 1874  & 217  & 43 & 4104 &  660 & 522  \\
 \hline
  \end{tabular}
\caption{Number of LVS CY geometries suitable for poly-instanton inflation. Here $W$ denotes a generic Wilson divisors, while $W_\Pi$ a Wilson divisor with $\Pi=0$.}
\label{tab_GwilsonLVS}
\end{table}

As a side remark, let us recall that for having poly-instanton corrections to the superpotential one needs to find a Wilson divisor $W$ with $h^{2,0}(W) = 0$ and $h^{0,0}(W)= h^{1,0} (W)=  h^{1,0}_+(W)=1$ for some specific choice of involution, without any restriction on $h^{1,1}(W)$ \cite{Blumenhagen:2012kz}. On these lines, a different type of `Wilson' divisor suitable for poly-instanton corrections has been presented in \cite{Lust:2013kt}, which has $h^{1,1}(W) = 4$ instead of 2, and so it has a non-vanishing $\Pi$. As we will discuss in a moment, this means that any poly-instanton inflation model developed with such an example would not have leading order protection against higher-derivative $F^4$ corrections for the inflaton direction $\tau_w$. Tab. \ref{tab_GwilsonLVS} and \ref{tab_Wilsonmismatch-Gstar} show the existence of several Wilson divisors which fail to have vanishing $\Pi$ since they have $h^{1,1}(W) \neq 2$.

\noindent
 \begin{table}[H]
  \centering
 \begin{tabular}{|c||c|c||c|c|c|c||c|c|c|c|}
\hline
 $h^{1,1}$ & Poly$^*$  & Geom$^*$  & at least  & single & two & three  & at least  & single & two & three      \\
&  &  & one $W$ & $W$ & $W$  & $W$  & one $W_\Pi$ & $W_\Pi$ & $W_\Pi$  & $W_\Pi$      \\
 \hline
 1 & 5 & 5 & 0 & 0  & 0 & 0 & 0  & 0 & 0 & 0  \\
 2 & 36 & 39 &  0 & 0  & 0 & 0 & 0  & 0 & 0 & 0  \\
 3 & 243 & 305 & 19 &  19 & 0  & 0  & 19 & 19  & 0  & 0    \\
 4 & 1185 & 2000 & 229 & 221  & 8 & 0 & 210 & 202 & 8  & 0    \\
 5 & 4897 &13494 & 2134  & 1874  & 217 & 43 &  1764 & 1599  &  154 & 11 \\
 \hline
  \end{tabular}
\caption{CY geometries with Wilson divisors $W$ and vanishing $\Pi$ Wilson divisors $W_\Pi$ without demanding a diagonal del Pezzo divisor.}
\label{tab_Wilsonmismatch-Gstar}
\end{table}

\subsection{Comments on $F^4$ corrections}

The higher-derivative $F^4$ corrections to the potential of poly-instanton inflation can be written as:
\bea
\label{VF4simpl}
V_{F^4} &= & \frac{\gamma}{{\cal V}^4} \, \left(\Pi_b\, t^b  + \Pi_s \, t^s + \, \Pi_w \, t^w \right)\, \\
&  = & \frac{\gamma}{{\cal V}^4} \, \biggl[\Pi_b\,  \left(\frac{6\,}{k_{bbb}}\right)^{1/3}\, \left({\cal V} +\beta_s \, \, \tau_{s}^{3/2} + \beta_s\, \, (\tau_{s}+\tau_w)^{3/2} \right)^{1/3}  \nonumber\\
&-& \Pi_s \, \sqrt{\frac{2\,\tau_{s}}{k_{sss}}}  - \, \Pi_w \, \sqrt{\frac{2\,(\tau_s+\tau_w)}{k_{sss}}} \biggr] \,, \nonumber
\eea
where we have used:
\be
t^b = \, \sqrt{\frac{2\, \tau_b}{k_{bbb}}}, \quad t^s = - \, \sqrt{\frac{2}{k_{sss}}} \, \left(\sqrt{\tau_{s}} + \sqrt{\tau_s + \tau_w}\right)\,, \quad t^w = -\, \sqrt{\frac{2}{k_{sss}}} \, \sqrt{\tau_s + \tau_{w}}\,.
\ee
Now we know that for our Wilson divisor case, $\Pi_w = 0$, and so the last term in (\ref{VF4simpl}) automatically vanishes. This gives at least a leading order protection for the potential of the inflaton modulus $\tau_w$ after stabilising the ${\cal V}$ and $\tau_s$ moduli through LVS. However the $\tau_w$-dependent term proportional to $\Pi_b$ would still induce a subleading inflaton-dependent correction that scales as ${\cal V}^{-14/3}$. As compared to the LVS potential, this $F^4$ correction is suppressed by a ${\cal V}^{-5/3}$ factor which for $\vo\gg 1$ should be small enough to preserve the predictions of poly-instanton inflation studied in \cite{Blumenhagen:2012ue,Gao:2013hn, Gao:2014fva}. Interestingly, we have found that $F^4$ corrections to poly-instanton inflation can be topologically tamed, unlike the case of blow-up inflation. In fact, the topological taming of higher derivative corrections to blow-up inflation would require the inflaton to be the volume of a diagonal dP$_3$ divisor which, according to the conjecture formulated in \cite{Cicoli:2021dhg}, is however very unlikely to exist in CY threefolds from the KS database.

\section{Summary and conclusions}
\label{sec_conclusions}

In this article we presented a general discussion of the quantitative effect of higher derivative $F^4$ corrections to the scalar potential of type IIB flux compactifications. In particular, we discussed the topological taming of these corrections which \emph{a priori} might appear to have an important impact on well-established LVS models of inflation such as blow-up inflation, fibre inflation and poly-instanton inflation. 

These $F^4$ corrections are not captured by the two-derivative approach where the scalar potential is computed from the K\"ahler potential and the superpotential, since they directly arise from the dimensional reduction of 10D higher derivative terms. In addition, such a contribution to the effective 4D scalar potential turns out to be directly proportional to topological quantities, $\Pi_i$, which are defined in terms of the second Chern class of the CY threefold and the (1,1)-form dual to a given divisor $D_i$. The fact that these higher derivative $F^4$ terms have topological coefficients has allowed us to perform a detailed classification of all possible divisor topologies with $\Pi=0$ that would lead to a topological taming of these corrections. In particular, we have found that the divisors with vanishing $\Pi$ satisfy $\chi(D) = 6 \chi_h(D)$ which is also equivalent to the following relation among their Hodge numbers: $h^{1,1}(D) = 4 \, h^{0,0}(D) - 2\, h^{1,0}(D) + 4 \, h^{2,0}(D)$. In order to illustrate our classification, we presented some concrete topologies with $\Pi=0$ which are already familiar in the literature. These are, for example, the 4-torus ${\mathbb T}^4$, the del Pezzo surface of degree-6 dP$_3$, and the so-called `Wilson' divisor with $h^{1,1}(W) = 2$. 

In search of seeking for divisors of vanishing $\Pi$, we investigated all (coordinate) divisor topologies of the CY geometries arising from the 4D reflexive polytopes of the Kreuzer-Skarke database. This corresponds to scanning the Hodge numbers of around 140000 divisors corresponding to roughly 16000 distinct CY geometries with $1 \leq h^{1,1}(X)\leq 5$. In our detailed analysis, we have found only two types of divisors of vanishing $\Pi$: the dP$_3$ surface and the `Wilson' divisor with $h^{1,1}(W) = 2$. 

In addition to presenting the scanning results for classifying the divisors of vanishing $\Pi$, we have also presented a classification of CY geometries suitable to realise LVS moduli stabilisation and three different inflationary models, namely blow-up inflation, fibre inflation and poly-instanton inflation. Subsequently, we studied numerically the effect of $F^4$ corrections on these inflation models in the generic case where the inflaton is not a divisor with vanishing $\Pi$. In this regards, we performed a detailed analysis of the post-inflationary evolution to determine the exact number of efoldings of inflation to make contact with actual CMB data. When the coefficients of the $F^4$ corrections are non-zero, we found that they generically do not spoil the predictions for the main cosmological observables. A crucial help comes from the $(2\pi)^{-4}$ suppression factor present in (\ref{eq:suppresion}) which gives the coefficient of higher derivative corrections for the $h^{1,1}(X) =1$ case. However, we argued that this suppression factor should be universally present in all $F^4$ corrections of the kind presented in this work, even for cases with $h^{1,1}(X) >1$.

Let us finally mention that our detailed numerical analysis shows that all the three LVS inflationary models, namely blow-up inflation, fibre inflation and poly-instanton inflation, turn out to be robust and stable against higher derivative $\alpha'^3$ corrections, even for the cases when such effects are not completely absent thanks to appropriate divisor topologies in the underlying CY orientifold construction. In some cases, like in blow-up inflation, we have even found that such corrections can help to improve the agreement with CMB data of the prediction of the scalar spectral index. 

It is however important to stress that these are not the only corrections which can spoil the flatness of LVS inflationary potentials. To make these models more robust, one should study in detail the effect of additional corrections, like for example string loop corrections to the potential of blow-up and poly-instanton inflation. In this paper, we have assumed that these corrections can be made negligible by considering values of the string coupling which are small enough, or tiny flux-dependent coefficients. However, this assumption definitely needs a deeper analysis since in LVS the overall volume is exponentially dependent on the string coupling, and $\vo$ during inflation is fixed by the requirement of matching the observed value of the amplitude of the primordial density perturbations. Therefore taking very small values of $g_s$ to tame string loops might lead to a volume which is too large to match $A_s$. We leave this interesting analysis for future work.

\acknowledgments

FQ would like to thank Perimeter Institute for Theoretical Physics for hospitality during the late stages of this project. PS would like to thank the {\it Department of Science and Technology (DST), India} for the kind support.

\bibliographystyle{JHEP}
\bibliography{reference}

\end{document}